\documentclass[a4paper,11pt]{article}
\pdfoutput=1 

\usepackage{jinstpub} 

\usepackage{amssymb}
\usepackage{wasysym}
\usepackage{amsmath}
\usepackage{upgreek}
\usepackage{xspace}
\usepackage{xfrac}

\begin{document}

\newcommand{\dk}{\mathrm{d}k}
\newcommand{\dd}{\mathrm{d}}
\newcommand{\scoh}{\sigma_\mathrm{coh}}
\newcommand{\sinc}{\sigma_\mathrm{inc}}
\newcommand{\mdr}{\mu\!/\!\rho}
\newcommand{\Fsc}{$F_\mathrm{sc}$ }
\newcommand{\kev}{\mbox{ke\hspace{-0.09 em}V}\xspace}
\newcommand{\kv}{\mbox{k\hspace{-0.05 em}V}\xspace}

\title{Optimal X-ray filters for EDXRF analysis}

\author[a,1]{D. Maier,\note{Corresponding author.}}
\author[a]{O. Limousin,}
\author[a]{D. Renaud,}
\author[a]{F. Visticot}

\affiliation[a]{AIM, CEA, CNRS, Universit\'e Paris-Saclay, Universit\'e Paris Diderot,\\Sorbonne Paris Cit\'e; F-91191 Gif-sur-Yvette Cedex; France}

\emailAdd{daniel.maier@cea.fr}

\abstract{This work presents a semi-analytical approach to answer the question of optimal beam filtering in the case of EDXRF measurements with an X-ray tube. A collection of programs, called \textit{xfilter}, is presented that is capable to find the optimal filter material
and the optimal scattering angle for all possible combinations of trace elements and target materials. The tube voltage can be either set fixed or be another parameter to optimize.
The concepts of the calculations are introduced in a general manner and demonstrated with a specific example, the detection of gold K$_{\upalpha 1}$ XRF within human tissue. The comparison between the calculation results and an EDXRF measurement shows excellent agreement.}


\arxivnumber{1907.13552} 

\maketitle
\flushbottom

\section{Introduction}
\label{intro}
Energy dispersive X-ray fluorescence (EDXRF) is widely used for quantitative and qualitative element analysis.\,\cite{Beckhoff2007} EDXRF systems based on a triaxial geometry\,\cite{Standzenieks1979, Bisgard1981} or based on synchrotron radiation\,\cite{Knoechel1985} are used for more than 30 years now. Compared to an EDXRF system based on a direct X-ray tube irradiation, both methods show a superior sensitivity but their usage is restricted because of the low flux that comes with a triaxial geometry or the limited availability of a synchrotron beam.  In contrast, a conventional X-ray tube in combination with an appropriate filter is used in many applications where cost, portability, size, and availability is an issue.

\begin{figure}[ht]
   \centering
   \includegraphics[width = 0.95\linewidth]{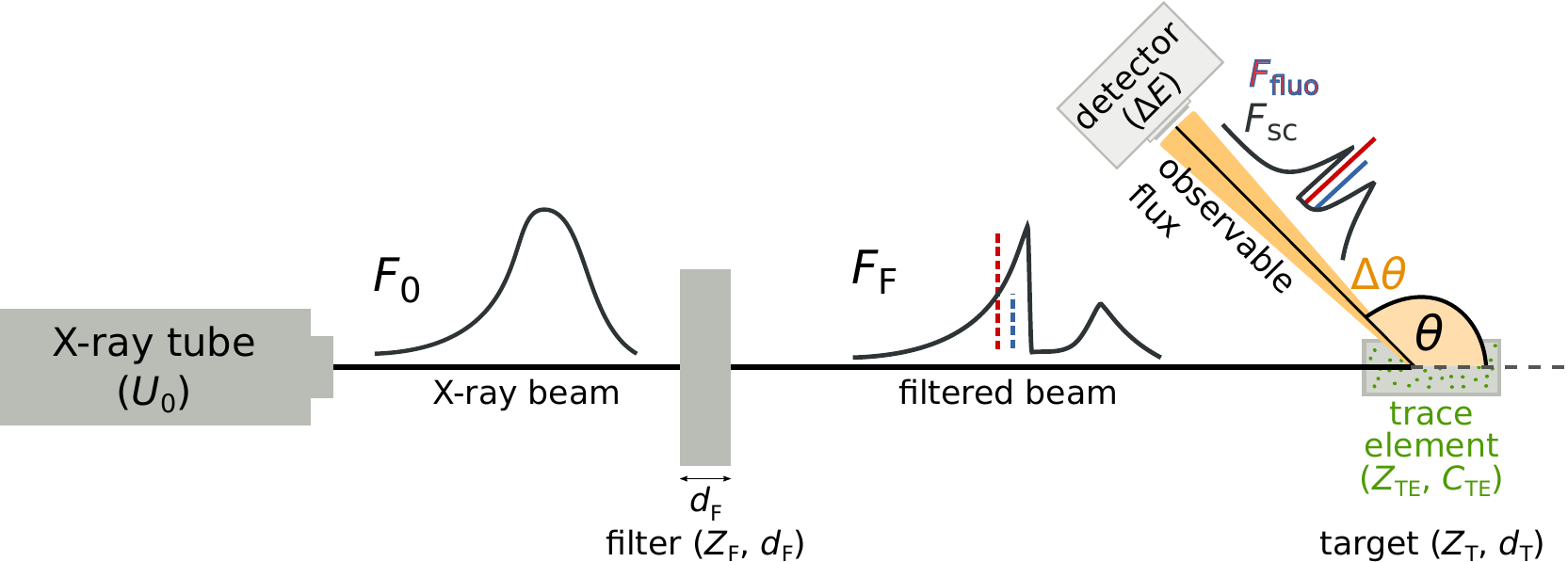}
   \caption{Principal geometry of an EDXRF setup using a beam filter. Even though the fluorescence radiation is not present in the filtered beam spectrum, it's position is indicated as dashed lines in order to illustrate the shift of the background radiation by scattering.}
   \label{fig:edxrf}
\end{figure}

Figure~\ref{fig:edxrf} shows the principal geometry of an EDXRF setup using a primary beam filter. The X-ray beam of an X-ray tube with flux $F_0$ is modified by a filter resulting in the filtered flux $F_\mathrm{F}$ that hits a target with atomic number $Z_\mathrm{T}$. The target contains trace elements with atomic number $Z_\mathrm{TE}$ in a typical concentration $C_\mathrm{TE}$ between several parts per million and several percent. A detector unit measures a part of the fluorescence radiation of the trace elements, with flux $F_\mathrm{fluo}$, as signal and a part of the scattered beam radiation, with flux $F_\mathrm{sc}$, as background component. The spectral and angular distribution of $F_\mathrm{fluo}$ and $F_\mathrm{sc}$ differ from each other allowing to set the experimental conditions in a way that optimizes the signal-to-background ratio for the X-ray fluorescence (XRF) detection. 

Using a filter for the primary X-ray beam in order to optimize the signal-to-background ratio for EDXRF measurements is reported in several publications\,\cite{Gilmore1968, Potts1986, Ogawa2010, Manohar2014, Pessanha2017}.
Gilmore et al.\,\cite{Gilmore1968} demonstrated an improved detectability of As in a hydrocarbon target using a 40\,$\upmu$m thick Zn filter and a tube voltage of $U_0 = 50\,$\kv. 
Potts et al.\,\cite{Potts1986} showed that a 12.5\,$\upmu$m Fe foil for $U_0 = 20\,$\kv can be used to suppress a disturbing Fe fluorescence line effectively allowing the detection of Cr, V, and Ba in an Fe-rich environment.
Ogawa et al.\,\cite{Ogawa2010} calculated and measured the lower detection limit of Pb in brass for Al, Ti, Ni, Zr, and Mo filters with thicknesses between 20 and 125\,$\upmu$m for $U_0 = 50\,$\kv. Within this range, they demonstrated that the detection limit decreases as the filter thickness increases.
A Monte Carlo study by Manohar et al.\,\cite{Manohar2014} compared a 1, 2, and 3\,mm thick Sn and a 2 and 3\,mm thick Pb filtration for the detection of Au in a PMMA target using $U_0 = 105\,$\kv. They concluded that the Sn filter works better than the Pb filter and that the thickness must be chosen as a trade-off between signal-to-dose value and duration of acquisition.
Pessanha et al.\,\cite{Pessanha2017} showed measured detection limits for up to 13 elements within 5 different target materials using $U_0=30\,$\kv for the following filter combinations: no filter, 250\,$\upmu$m Al, 25\,$\upmu$m Cu, 250\,$\upmu$m Al + 25\,$\upmu$m Cu. They concluded that for fluorescence energies $E_\mathrm{fluo} < 5$\,\kev no filtering is beneficial, while the use of Al and Cu filters can improve the detection limit for $5 < E_\mathrm{fluo}\,[\mathrm{\kev}] < 15$. This improvement is more pronounced for low-Z targets.
More examples of recent investigations on primary beam filtering for EDXRF analysis can be found in Pessanha et al.\,\cite{Pessanha2017} and the references therein.

Using an experimental approach clearly limits the number of free parameters to optimize EDXRF analysis. More specifically, the limitations are:
\begin{itemize}
\setlength\itemsep{0.0em}
\item The voltage of the X-ray tube $U_0$ and the scattering angle $\theta$ were considered to be fix within all of the cited works.
\item The number of filter materials and filter thicknesses are limited to a few examples and do not allow to conclude on the optimal filtering but only on the best filter within the test sample.
\item The number of target materials and trace elements are limited to the given examples.
\end{itemize}

A framework capable to optimize EDXRF measurements, i.e.~to find the optimal tube voltage, the optimal filter material and filter thickness, and the optimal scattering angle, for all possible combinations of trace elements and target materials is clearly missing. 

Our motivation to optimize EDXRF measurements is based on the project SATBOT which consists of a CdTe based, robotized, X-ray spectro-imaging camera\,\cite{Maier2018} that assists gold nano\-par\-ticle-enhanced radiotherapy\,\cite{Kuncic2018}. It aims to conduct near real-time dosimetry\,\cite{patent_EP}, quantification of administered nanoparticles\,\cite{patent_EP}, and 3D-XRF computed tomography\,\cite{Vienne2018}. Because of this background, the presented optimization framework is verified with measured data for $Z_\mathrm{TE}=79$~(Au) in a water phantom with an effective atomic number of $Z_\mathrm{T} \approx 7$. Despite this specific case,
the semi-analytical approach presented in this work aims to answer the question of optimal filtering solutions ($U_0$, $Z_\mathrm{F}$, $d_\mathrm{F}$, $\theta_\mathrm{opt}$) for any given test setup ($Z_\mathrm{T}$, $Z_\mathrm{TE}$) in an approximated manner. The approximation is made by applying a thin target approximation for the photon interactions in the target and, because of that, focusing on first order photon interactions only. In a second step, a parameter fine-tuning based on Monte Carlo simulations or measurements may be following this initial solution.

\section{Theory}
\subsection{Thin target approximation}
Given an attenuation coefficient $\mu_\mathrm{x}$ for a specific photon interaction $\mathrm{x}$, the probability density function ($\mathit{pdf}$) for an interaction within $z \pm \dd z/2$ is
\begin{equation}
   \label{eq:beer_pdf}
   \mathit{pdf}_\mathrm{\!x} = \mu_\mathrm{x} \cdot \exp{\!(-\mu_\mathrm{tot} \, z)} \cdot \mathrm{d}z
\end{equation}
with $\mu_\mathrm{tot}$ being the total attenuation coefficient.
Integrating Eq.\,(\ref{eq:beer_pdf}) over the detector thickness $d$ yields the probability of interaction
\begin{equation}
   \label{eq:tta0i}
   P_\mathrm{x} = \big(1 - \exp(-\mu_\mathrm{tot}\,d)\big) \cdot \frac{\mu_\mathrm{x}}{\mu_\mathrm{tot}}.
\end{equation}
The thin target approximation (TTA) uses the following approximation
\begin{equation}
   1-\exp{(-x)} \approx x \quad \textnormal{for } x \ll 1
\end{equation}
which implies a relative error less than 5\,\% for $x < 0.1$.
Using TTA and $d = \rho_{\!A}/\rho$, Eq.\,(\ref{eq:tta0i}) becomes
\begin{equation}
   \label{eq:tta1,2}
   P_\mathrm{x} \approx \mu_\mathrm{x}\!/\!\rho \cdot \rho_{\!A} \quad \textnormal{for } \mu_\mathrm{tot}\!/\!\rho \cdot \rho_{\!A} \ll 1
\end{equation}
i.e.~the interaction probability per area density $\rho_{\!A}$ is given by the cross section $\mu_\mathrm{x}\!/\!\rho$. An important consequence of Eq.\,(\ref{eq:tta1,2}) is that a specific photon interaction can be analyzed independently from the total cross section $\mu_\mathrm{tot}\!/\!\rho$.

\begin{figure}[tbh]
   \centering
   \includegraphics[width = \linewidth]{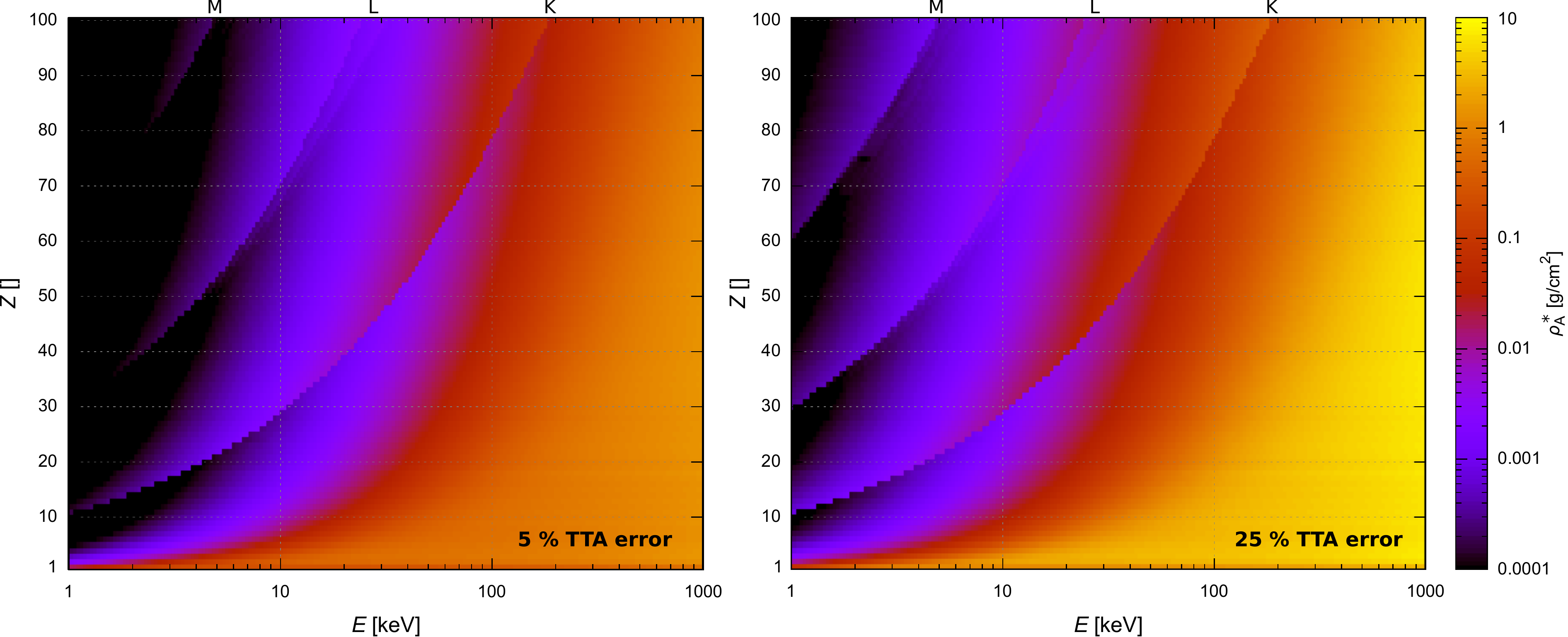}
   \caption{\textbf{Left:} critical area density $\rho_{\!A}^*$ that yields a 5\,\% TTA error for a photon with energy $E$ interacting in a target material with atomic number $Z$. \textbf{Right:} same for a 25\,\% TTA error. The K, L, and M-edges are indicated on top of the plots. Example: a photon interaction in water ($Z_\mathrm{eff} \approx 7$) with $E=100\,$keV yields a 5\,\% error for a 1\,cm thick target and a 25\,\% error for a 4\,cm thick target.}
   \label{fig:TTA_error}
\end{figure}

\noindent
Equation~(\ref{eq:tta1,2}) shows that the total cross section $\mu_\mathrm{tot}\!/\!\rho$ and the area density $\rho_{\!A}$ determine the applicability of TTA. Figure~\ref{fig:TTA_error} shows the critical area density $\rho_{\!A}^*$ that yields a TTA error of 5\,\% and 25\,\% for all elements and $1\,\mathrm{keV} \leq E \leq 1\,\mathrm{MeV}$. 

In the following calculations TTA is used for fluorescence generation and for target scattering. In the case of fluorescence generation this is not a severe limitation. Looking on the right side of the K-edge in Fig.\,\ref{fig:TTA_error} (left) shows that $\rho_{\!A}^* > 1\,$mg/cm$^2$ for $Z>30$. The area density of trace elements is in most cases lower than this value. Only in cases of low-Z trace elements or L-line fluorescence measurements TTA might be invalid for fluorescence generation. 

For target scattering, TTA is a more severe limitation. Applying TTA with a maximal error of 5\,\% requires for $Z\leq 20$ and $E\geq100\,$keV \mbox{$\rho_{\!A} \mbox{\text{\footnotesize$\, \lesssim \,$}} 1\,\mathrm{g}/\mathrm{cm}^2$} for the target material. For lower energies and higher atomic numbers the validity of TTA must be evaluated for the used area density of the target material.
The reason why the following calculations are still using TTA is
\begin{itemize}
\setlength\itemsep{0.0em}
\item the linearity of Eq.\,(\ref{eq:tta1,2}): using TTA for fluorescence generation and for target scattering means that we can assume a linear relation between the signal and the area density of the trace elements $\rho_{\!A_\mathrm{TE}}$ and between the background and the area density of the target material~$\rho_{\!A_\mathrm{T}}$. This linear relation allows us to state the signal per area density of trace elements and the background per area density of target material; these results can be directly scaled to different $\rho_{\!A_\mathrm{TE}}$ and $\rho_{\!A_\mathrm{T}}$ for other experiments.
\item non-TTA conditions also enforce more complex considerations like multiple scattering which are intensive in terms of computation and which depend on the geometry of the objects. For these cases individualized Monte Carlo simulations are mandatory.
\end{itemize}
Nevertheless, it is important to note that properly collimated beam and detector setups as well as a proper preparation of the target allow to use TTA in many cases.

\subsection{Photon interactions}
In the typical energy range of the primary X-ray beam (tens to hundreds of \kev), the dominant photon interactions are coherent scattering, incoherent scattering, and photoelectric absorption. We are limiting the discussion to unpolarized light as this is the case for X-ray tube radiation.

\begin{figure}[tbh]
   \centering
   \includegraphics[width = 0.95\linewidth]{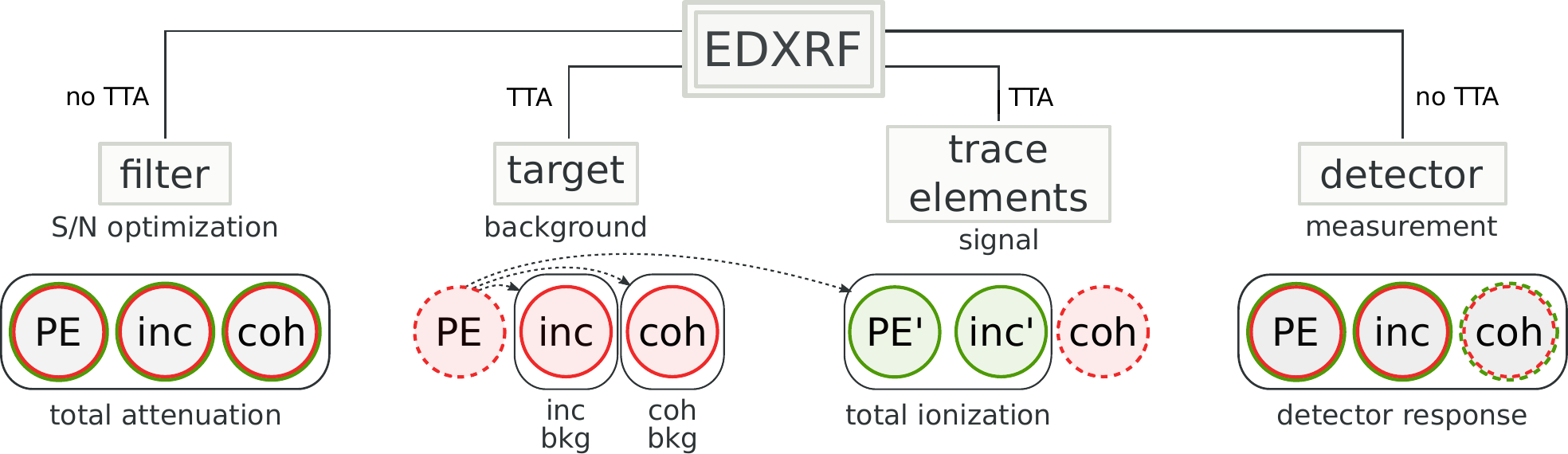}
   \caption{Photon interactions within an EDXRF setup and their treatment within the presented work: photoelectric effect (PE), incoherent scattering (inc), and coherent scattering (coh). PE' and inc' indicate that only a fraction of the total interaction is of relevance. The colors red and green indicate background (bkg) and signal contributions, respectively. Dashed lines indicate interactions that are ignored.}
   \label{fig:analysis_approach}
\end{figure}

The following considerations focus on a photon-matter interaction analysis of the beam filter, the target material, and the trace elements. In addition, an accurate detector response model is required for the background and signal calculation; see Fig.\,\ref{fig:analysis_approach} for a schematic summary of the used photon interactions in the different parts of the calculation.
\begin{itemize}
\setlength\itemsep{0.0em}
\item Within the beam filter the total attenuation comprises the combination of photoelectric absorption, incoherent scattering, and coherent scattering and is expressed via a material and energy dependent total mass attenuation coefficient $\mu\!/\!\rho(E, Z)$. The attenuation of the filter affects the signal and the background of the EDXRF measurement.
\item Within the target coherent and incoherent scattering are generating the background radiation.
Photoelectric absorption within the target modifies the background as only those photons can contribute to the background that are scattered into the direction of the detector without being absorbed on their way to the detector.
This contribution is neglected as the photoelectric absorption of the background is the same as the photoelectric absorption of the signal at the signal energy $E_\mathrm{fluo}$ so that the signal-to-background ratio is unchanged by photoelectric absorption. If the detector response shows a strong Comptonization component this argument is no longer true, as photons with $E \gg E_\mathrm{fluo}$, i.e.~photons which are less absorbed by photoelectric absorption, can contribute to the background at $E_\mathrm{fluo}$. A Monte Carlo simulation must be conducted in this case.
\item Within the trace elements, photoelectric absorption and incoherent scattering are partially relevant for the signal generation. Their contribution is defined by those interactions that ionize the atomic shell of the fluorescence line of interest. A fraction of these ionized atoms, which is expressed by the combination of the line strength and the fluorescence yield, constitute the signal. Background contributions of coherent and incoherent scattering within the trace elements can be ignored because of the low concentration of trace elements.
\item The detector response describes the redistribution of the photon energy within the detector material. The used model takes into account: photoelectric absorption, escape of K lines, Comptonization within the detector, and the energy resolution of the detector system.
\end{itemize}

In summary, five applications of photon matter interactions must be considered: the total filter attenuation, incoherent and coherent scattering from the target towards the detector, the fluorescence generation, and the detector response.

\subsubsection{Total filter attenuation}
An incoming flux $F_0$ that passes through a medium with area density $\rho_{\mathrm{\!A}}$, and mass absorption coefficient $\mdr$ is reduced to the filtered flux~$F_\mathrm{F}$ according to
\begin{equation}
   \label{eq:filter_abs}
   F_\mathrm{F} = F_0 \cdot \mathrm{e}^{-\mdr\, \cdot \, \rho_{\mathrm{\!A}}}.
\end{equation}
Values for $\mdr(E,Z)$ are based on lookup tables\,\cite{xcom-online} in combination with log-log cubic spline interpolations conducted by \textit{xdata}\,\cite{xdata}.

\subsubsection{Fluorescence generation}
\label{sec:fluo_gen}
\begin{figure}[ht]
   \centering
   \includegraphics[width = \linewidth]{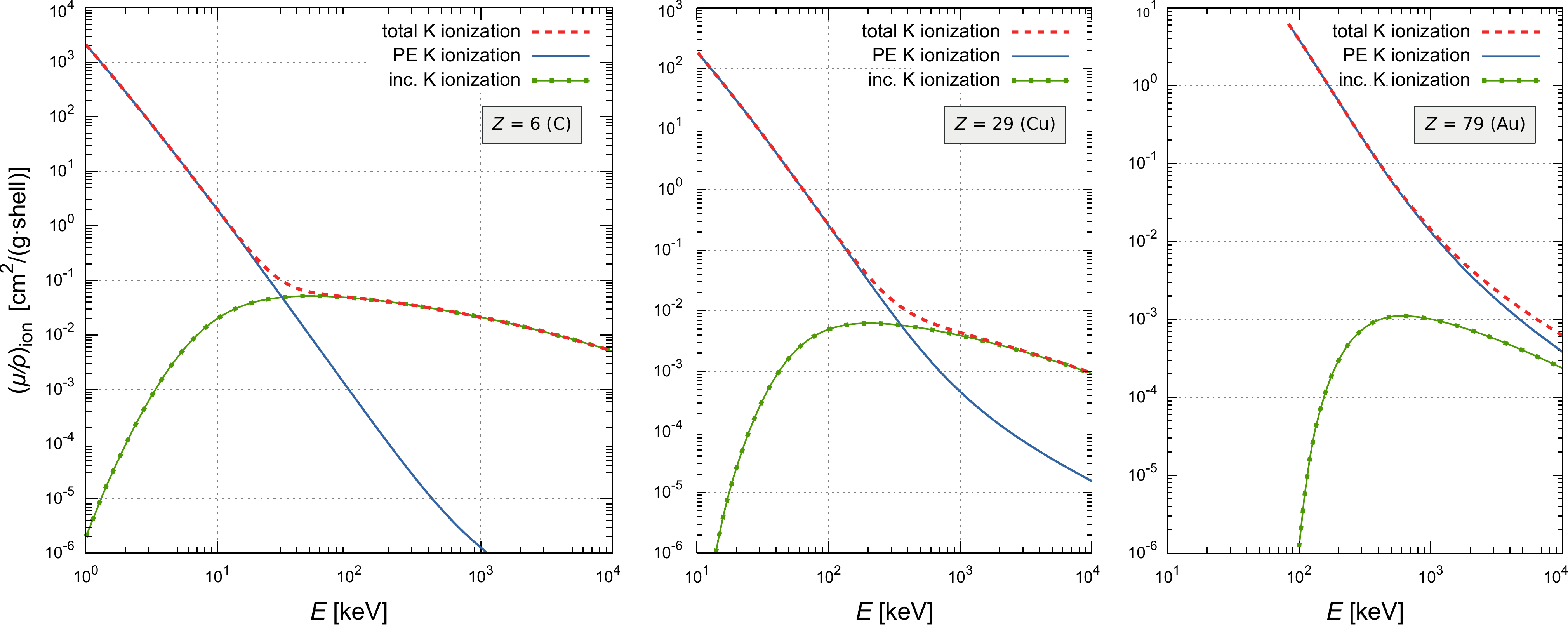}
   \caption{K-shell ionization cross sections for the materials C, Cu, and Au. The total K-shell ionization cross section is the sum of K-shell ionizations caused by photoelectric effect (PE) and by incoherent scattering (inc). Data: \textit{xdata}\,\cite{xdata}.}
   \label{fig:ion}
\end{figure}
The fluorescence emission depends on the ionization rate $\dot{I}$, the fluorescence yield $Y_s$ for the shell $s$ under consideration, the branch ratio $R_l$ for the line $l$ under consideration via
\begin{equation}
  \label{eq:sig_1}
   F_{\mathrm{fluo},s,l} = \frac{1}{4\pi} \, \dot{I} \cdot Y_s \cdot R_l \qquad \textnormal{in units } \left[\frac{1}{\textnormal{s$\cdot$sr$\cdot$g}}\right]\!\!.
\end{equation}
The ionization rate caused by an incoming flux $F_\mathrm{F}$ can be calculated via
\begin{equation}
   \label{eq:sig_2}
   \dot{I} = \frac{1}{\rho_{\!A}} \, \int\limits_0^{\infty}\!F_\mathrm{F}(E) \cdot P_\mathrm{ion}(E) \cdot \mathrm{d} E \qquad \textnormal{in units } \left[\frac{1}{\textnormal{s$\cdot$g}}\right]
\end{equation}
with
\begin{align}
  P_\mathrm{ion} &= \big(1-\exp(-(\mu\!/\!\rho)_\mathrm{tot} \cdot \rho_{\!A})\big) \cdot \frac{\mu_\mathrm{ion}}{\mu_\mathrm{tot}}\\
  \label{eq:p_ion}
                 &\approx (\mu\!/\!\rho)_\mathrm{ion} \cdot \rho_{\!A}
\end{align}
where Eq.\,(\ref{eq:p_ion}) uses TTA.
See Fig.\,\ref{fig:ion} for the total K-shell ionization cross section $(\mu\!/\!\rho)_\mathrm{ion}$ for different materials.

\subsubsection{Coherent background}
\label{sec:coh}
Coherent scattering is modeled with the Thomson cross section $\sigma_\mathrm{Th}$ in combination with the atomic form factor $F$ according to 
\begin{align}
   \frac{\sigma_\mathrm{Th}}{\dd\varOmega} &= \frac{r_\mathrm{e}^2}{2} \cdot \left(1 + \cos^2(\theta)\right)\\
\label{eq:coh}
   \frac{\dd \scoh}{\dd \varOmega} &= \frac{\dd \sigma_\mathrm{Th}}{\dd \varOmega} \cdot F^2
\end{align}
where $r_\mathrm{e}$ is the classical electron radius, $\theta$ the scattering angle, and $\sigma_\mathrm{coh}$ the cross section for coherent scattering. The atomic form factor $F$ accounts for the scattering on the electron cloud of an atom. Values for $F$ depend on the atomic number $Z$, the energy of the radiation $E_0$, and the scattering angle $\theta$ and are interpolated from tabulated values\,\cite{Hubbell1975}.

\subsubsection{Incoherent background}
\label{sec:inc}
Incoherent scattering is modeled with the Klein-Nishina cross section $\sigma_\mathrm{KN}$ in combination with the incoherent scattering function $S$ according to 
\begin{align}
   \label{eq:classicCompt}
   \frac{\dd \sigma_\mathrm{KN}}{\dd \varOmega} &= \frac{r_\mathrm{e}^2}{2}  \cdot \left(\frac{E'_\mathrm{c}}{E_0}\right)^{\!\!2} \cdot \left[\frac{E'_\mathrm{c}}{E_0} + \frac{E_0}{E'_\mathrm{c}} - \sin^2(\theta) \right]\\
   \label{eq:classicCompt2}
   \frac{\dd \sinc}{\dd \varOmega} &= \frac{\dd \sigma_\mathrm{KN}}{\dd \varOmega} \cdot S
\end{align}
with the following relation between the initial energy $E_0$ and the Compton scattered energy $E'_\mathrm{c}$ of the photon
\begin{equation}
\label{eq:kk0}
   E'_\mathrm{c} = \frac{E_0}{1 + \frac{E_0}{mc^2} \big(1 - \cos(\theta)\big)}
\end{equation}
with $m$ being the rest mass of the electron and $c$ is the speed of light.
The factor $S$ accounts for the fact that $\sigma_\mathrm{KN}$ is defined for scattering on a single free electron while for the interaction with atoms all bound electrons must be considered. Values for $S$ depend on the atomic number $Z$, the energy of the incident radiation $E_0$, and the scattering angle $\theta$ and are interpolated from tabulated values\,\cite{Hubbell1975}.

\subsubsection{Doppler broadening}
Another subtlety arises from the fact that the electrons in an atom are not at rest but have different momenta. The distribution of the electron momentum is expressed by the \textit{Compton profile} and results in a distribution of the scattered energy\footnote{The subscript \textit{c} used in Eq.\,(\ref{eq:kk0}) indicates a one-to-one relation between $E'$ and $E_0$ for a given scattering angle; no subscript is used for $E'$ in the case of Doppler broadening.} $E'$ for a fixed scattering angle $\theta$. The \textit{Doppler broadening} can be calculated according to the double-differential cross section (DDCS) derived by Ribberfors\,\cite{Ribberfors1975}
\begin{equation}
   \label{eq:DDCS}
   \frac{\dd^2\sigma_\mathrm{inc}}{\dd\varOmega\,\dd E'} = \frac{m r_\mathrm{e}^2}{2 E_0} \left(\!\frac{E'_\mathrm{c}}{E_0} + \frac{E_0}{E'_\mathrm{c}} - \sin^2(\theta) \! \right)  \frac{E'}{\sqrt{E_0^2 + E'^2 - 2 E_0 E' \cos(\theta)}} J_n(p_z)
\end{equation}
where $J_n(p_z)$ is the Compton profile for the $n^\mathrm{th}$ sub-shell of the target atom. Values for $J_n$ are taken from Biggs et al.\,\cite{Biggs1975}. The projection of the electron's pre-collision momentum on the momentum-transfer vector of the X-ray photon is\,\cite{Williams1977}
\begin{equation}
   \label{eq:pz}
   p_z = -mc \frac{E_0 - E' - E_0 E' \big(1-\cos(\theta) \big) / (mc^2)}{\sqrt{E_0^2 + E'^2 - 2 E_0 E' \cos(\theta)}}.
\end{equation}

\begin{figure}[t]
   \centering
   \includegraphics[width = \linewidth]{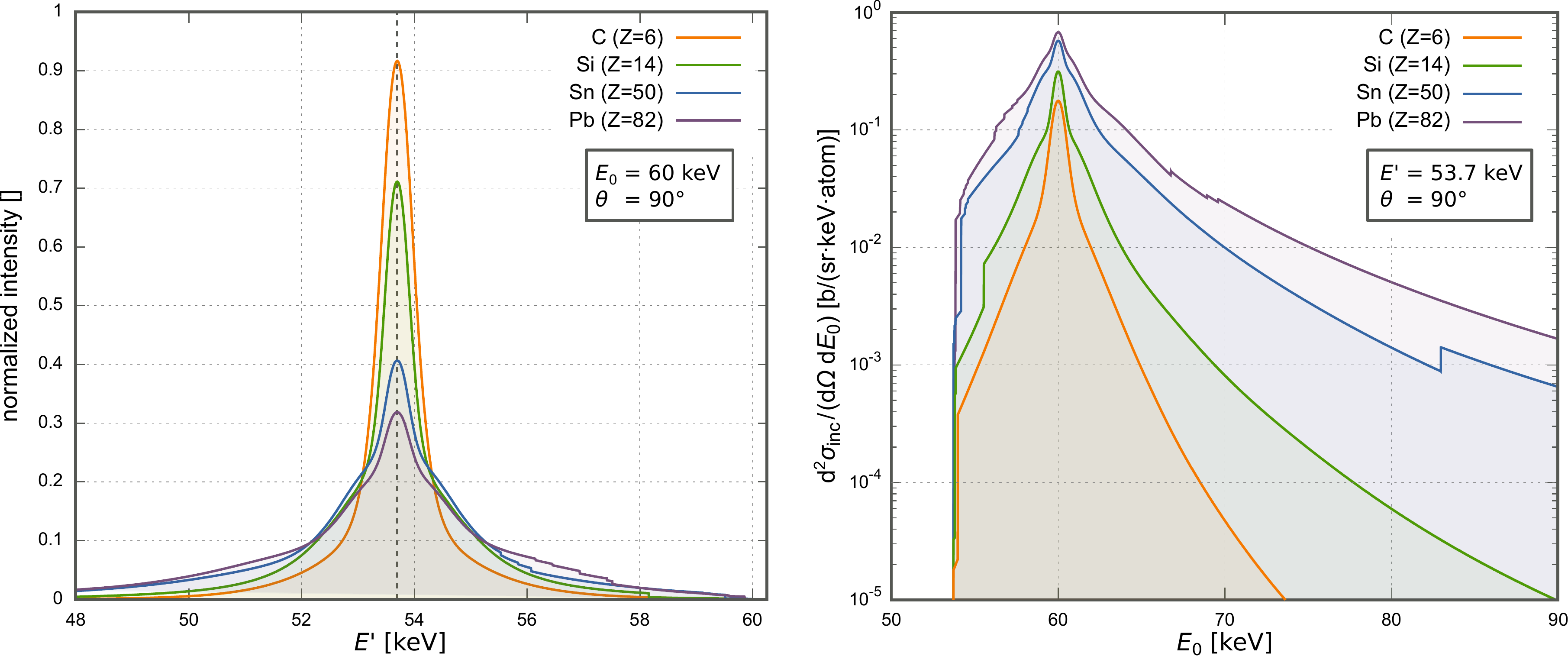}
   \caption{\textbf{Left:} Doppler broadening for $E_0 = 60\,$\kev at $\theta = 90^\circ$ for different scattering materials. The energy according to the Compton formula of Eq.\,(\ref{eq:kk0}), $E' = E'_c \approx 53.7$\,\kev, is shown as dashed line. \textbf{Right:} DDCS $\mathrm{d}^2\sigma_\mathrm{inc}/(\mathrm{d}\varOmega \, \mathrm{d} E_0)$ for $E' = 53.7$\,\kev. Data: \textit{xdata}\,\cite{xdata}.}
   \label{fig:doppler}
\end{figure}

Equation (\ref{eq:DDCS}) in combination with Eq.\,(\ref{eq:pz}) can be used to calculate the probability density function of the scattered photon energy.\,\cite{Ordonez1997} Figure \ref{fig:doppler} (left) shows Doppler broadening caused by different target atoms for a 90$^\circ$ scattering of photons with $E_0 = 60\,$\kev which is a representative energy for high-Z fluorescent materials. It is evident that Doppler broadening is an important effect to consider for high-resolution detector systems especially for scattering in high-Z target materials. An uncertainty for the Doppler broadening is difficult to state because the distribution is non-Gaussian and asymmetric. In addition, the activation of different atomic shells causes step-like structures.

Another important consequence of Eq.\,(\ref{eq:DDCS}) is that all photons with an energy that exceeds a specific energy of interest $E^*$, i.e.~$E_0 > E^*$, have a chance to be scattered to $E'=E^*$\!. 

While Eq.\,(\ref{eq:DDCS}) shows the distribution of scattered energies $E'$ for a given incident energy~$E_0$ and scattering angle $\theta$, a reversed interpretation of the DDCS shows the distribution of initial energies~$E_0$ for a given scattered energy~$E'$ and scattering angle~$\theta$, see Fig.\,\ref{fig:doppler}~(right). This distribution can be used to compute the scattering of the beam spectrum into a specific energy region of interest~$E' \pm \Delta E'\!/2$, see Sect.\,\ref{sec:bkg} and Eq.\,(\ref{eq:inc}).

\section{Implementation}
\label{sec:calc}
The implementation of the filter optimization is shown step by step, starting with the X-ray generation, the effect of filtering, up to the final signal and the background calculation. All these steps are demonstrated for one specific example, the detection of gold K$_{\upalpha 1}$ fluorescence in a carbon target. But, the calculations can be redone with the program \textit{xfilter}\,\cite{xfilter} for all K-fluorescence lines\footnote{The integration of the dominant L-lines is foreseen for the next version of \textit{xfilter}.} of any trace element in any target material. 

The calculations for the signal and background are using TTA, so that the signal and the background intensity are proportional to the area density~$\rho_{\!A}$ of the trace elements and the target material, respectively. 
All relevant quantities are defined as follows:
\begin{itemize}
\setlength\itemsep{-0.25em}
\item initial spectral flux density:\hspace{3.2mm} $\mathit{F}_0, \textnormal{in units}\, \big[{\mathrm{ph}}/({\kev\textnormal{$\cdot$s$\cdot$cm}^2})\big]$
\item filtered spectral flux density:\; $\mathit{F_\mathrm{F}}, \textnormal{in units}\, \big[{\mathrm{ph}}/({\kev\textnormal{$\cdot$s$\cdot$cm}^2})\big]$
\item signal intensity:\;\hspace{19.1mm} $\mathit{sig}, \textnormal{in units}\, \big[\mathrm{ph}/(\kev\textnormal{$\cdot$s$\cdot$g$\cdot$sr})\big]$
\item background intensity:\; $\hspace{9.3mm}\mathit{bkg}, \textnormal{in units}\, \big[\mathrm{ph}/(\kev \textnormal{$\cdot$s$\cdot$g$\cdot$sr})\big]\!.$
\end{itemize}
Be aware that the signal intensity is stated per gram of trace element while the background intensity is stated per gram of target material.
 
\subsection{X-ray generator}
\label{sec:xgen}
\begin{figure}[ht]
   \centering
   \includegraphics[width = \linewidth]{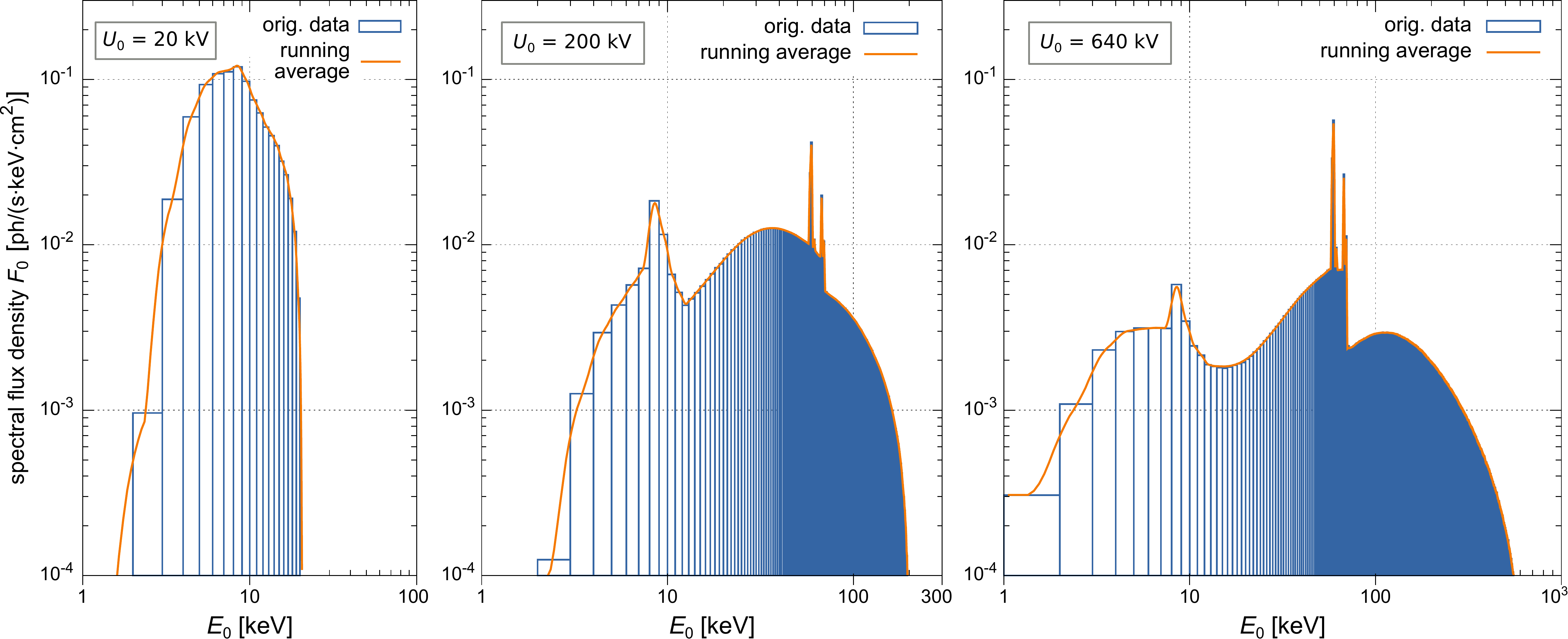}
   \caption{Normalized spectral flux density $F_0$. The graphs show the original data\,\cite{Hernandez_2014} (blue blocks) and the re-binned flux (orange line). The L-lines cannot be resolved because of the initial binning of 1\,keV.}
   \label{fig:tube_spec_all}
\end{figure}
All X-ray tube spectra are based on a Monte Carlo simulation\,\cite{Hernandez_2014} for tungsten anode spectra with acceleration voltages between $U_0 = 20\,\mathrm{\kv}$ and $U_0= 640\,\mathrm{\kv}$ and an intrinsic filtering of 0.8\,mm~Be. The original data are binned with a constant bin size of 1\,\kev. All data were re-binned to 0.1\,\kev and slightly smoothed with a two-step running average of window size $\pm 0.4\,$\kev and $\pm 0.2\,$\kev. Finally, all re-binned spectra were normalized in the following way
\begin{equation}
   \label{eq:norm}
   \int\limits_0^{\infty}\! F_0(E_0) \cdot \mathrm{d} E_0 = 1.
\end{equation}
The normalization is motivated by the idea that signal and background are both only originating from the tube flux and therefore, the signal-to-background ratio is independent of the tube flux but only dependent on the spectral shape of the X-ray beam.
Figure~\ref{fig:tube_spec_all} shows three illustrative spectra for $U_0 = 20\,$\kv, 200\,\kv, and 640\,\kv.

\subsection{Filter absorption}
\label{sec:filter}
\begin{figure}[htb]
   \centering
   \includegraphics[width = \linewidth]{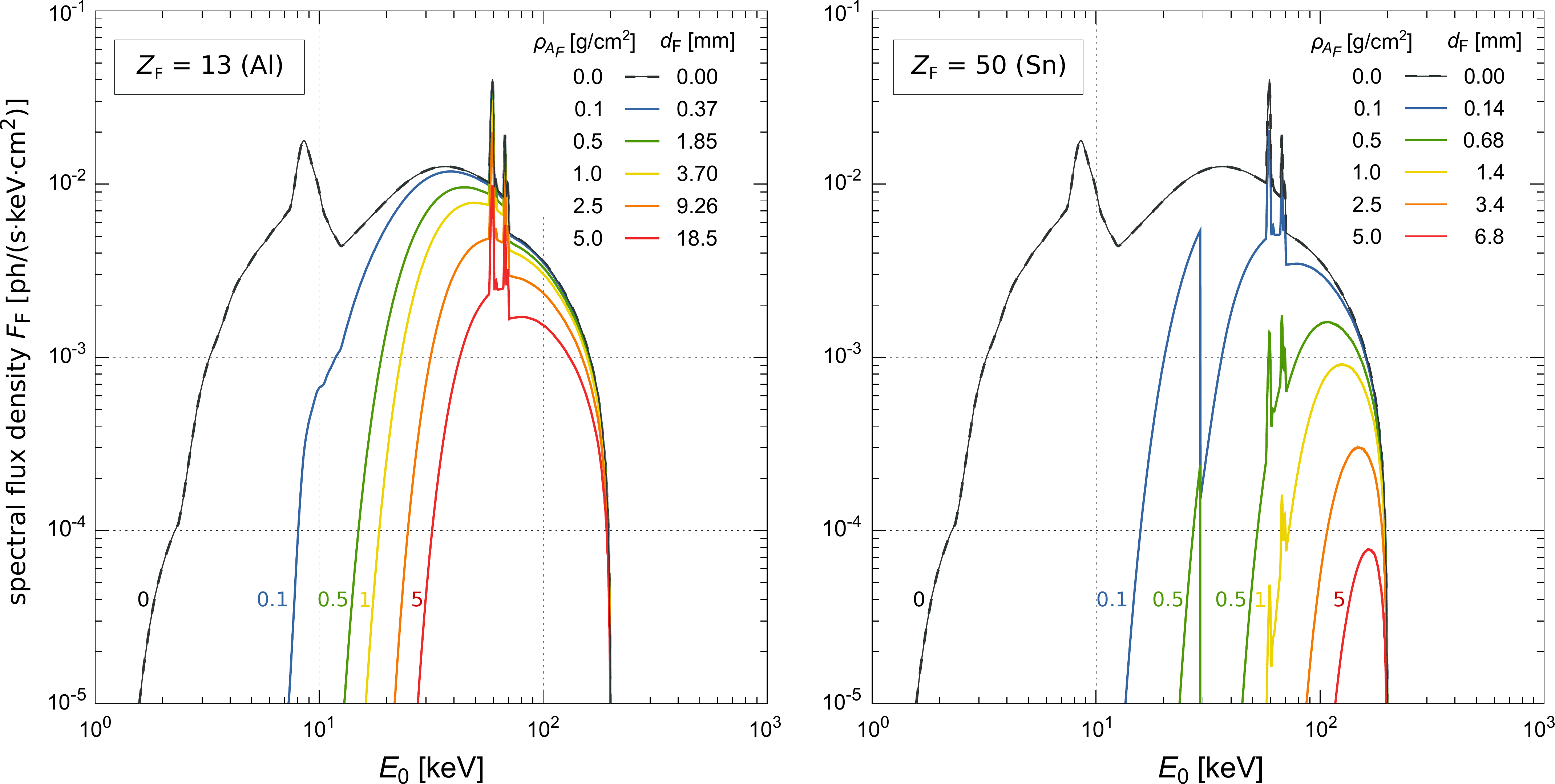}
   \caption{Spectral flux density $F_\mathrm{F}$ of the filtered beam for different area densities $\rho_{\!A_\mathrm{F}}$ for an aluminium, and tin filter. The corresponding filter thicknesses $d_\mathrm{F}$ are also indicated. The K-edge of Sn (29.2\,\kev) is clearly visible.}
   \label{fig:filter}
\end{figure}

The filter absorption is calculated according to Eq.\,(\ref{eq:filter_abs}). Figure~\ref{fig:filter} shows filtered spectra for two different filter materials ($Z_\mathrm{F} = 13$ and 50) and for different filter thicknesses $d_\mathrm{F}$ for an X-ray beam with $U_0 = 200\,$\kv. The sharp K-edge for the tin filter is clearly visible.

\subsection{Signal generation}
\label{sec:sig}
As described in Sect.\,\ref{sec:fluo_gen}, the signal intensity is calculated as the radiance of the fluorescence line of the trace element per area density~$\rho_{\!A_\mathrm{TE}}$. Combining Eq.\,(\ref{eq:sig_1}), (\ref{eq:sig_2}), and (\ref{eq:p_ion}) yields
\begin{align}
   F_{\mathrm{fluo},s,l} &\approx \frac{Y_s \cdot R_l}{4\pi}\int\limits_0^{\infty}\!F_\mathrm{F}(E_0)\cdot(\mu\!/\!\rho)_\mathrm{ion}(E_0)\cdot \mathrm{d}E_0\\
   \label{eq:signal}
     &\approx \frac{Y_s \cdot R_l}{4\pi} \sum\limits_{E_0=E_\mathrm{K}}^{E_\mathrm{max}} F_\mathrm{F}(E_0) \cdot (\mu\!/\!\rho)_\mathrm{ion}(E_0) \cdot \Delta E_0\\
   \label{eq:sig}
   \mathit{sig} &:= F_{\mathrm{fluo},s,l}/\Delta E_0 \quad \textnormal{in } \left[\frac{\mathrm{ph}}{\kev\textnormal{$\cdot$s$\cdot$g$\cdot$sr}}\right]\!\!.
\end{align}
\begin{figure}[t]
   \centering
   \includegraphics[width = \linewidth]{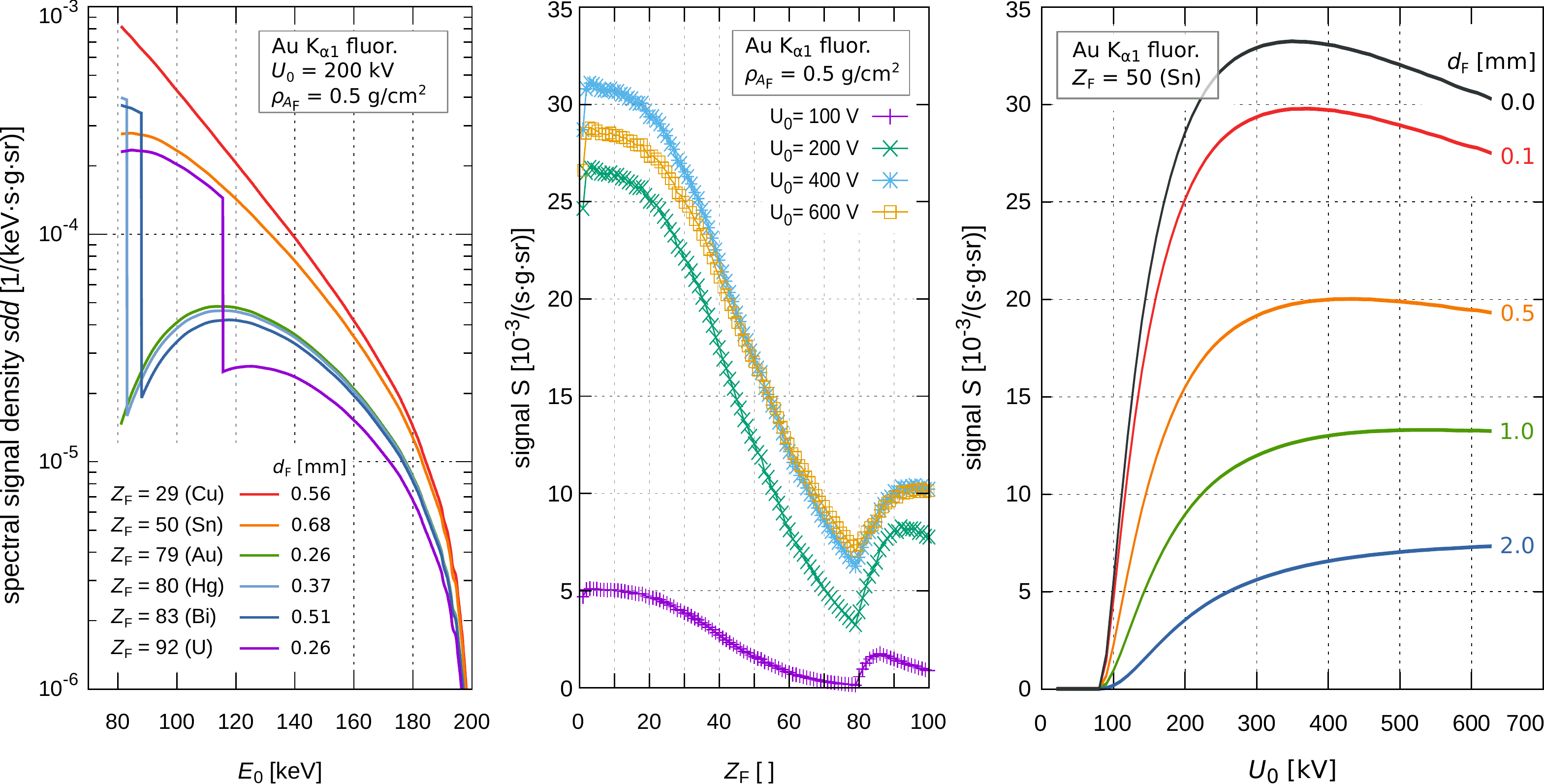}
   \caption[Signal generation for Au K$_{\upalpha 1}$]{Signal generation for Au K$_{\upalpha 1}$. \textbf{Left:} the spectral signal density states how much of the signal is produced by the impinging energy $E_0$. All filter thicknesses $d_\mathrm{F}$ are chosen according to $\rho_\mathrm{\!A}=0.5\,$g/cm$^2$. \textbf{Center:} integral signal value $S$ according to Eq.\,(\ref{eq:sig_s}) for all filter materials for different tube voltages $U_0$. \textbf{Right:} the signal $S$ shown for one filter with different thicknesses $d_\mathrm{F}$ as a function of the tube voltage $U_0$. The thicker the filter is chosen, the larger is the optimal tube voltage.}
   \label{fig:signal}
\end{figure}Equation~(\ref{eq:signal}) is calculated for an energy step size of $\Delta E_0 = 0.1\,$\kev between the K-edge $E_\mathrm{K}$ of the trace element and the maximal energy $E_\mathrm{max}$ defined by the used tube voltage $U_0$ and the charge of an electron $q_\mathrm{e}$ via 
\begin{equation}
   \label{eq:erg_max}
   E_\mathrm{max} = q_\mathrm{e} \cdot U_0.
\end{equation}
Equation~(\ref{eq:sig}) distributes the line emission over one bin width of size $\Delta E_0$ to get a spectral intensity. Figure~\ref{fig:signal} (left) shows the spectral signal density $\mathit{ssd}$
\begin{equation}
   \mathit{ssd}(E_0) = \frac{Y_s \cdot R_l}{4\pi} \cdot F_\mathrm{F}(E_0) \cdot (\mu\!/\!\rho)_\mathrm{ion}(E_0)
\end{equation}
 as a function of the beam energy for different filter materials with a constant area density of $\rho_{\!A_\mathrm{F}} = 0.5\,$g/cm$^2$. Above the K-edge, the increased absorption of high-Z materials is obvious. The total signal $S$ can be calculated as integral of the spectral signal density
\begin{equation}
   \label{eq:sig_s}
   S = \int\limits_0^{E_{\mathrm{max}}} \! \mathit{ssd}(E^*) \cdot \mathrm{d} E^*
\end{equation}
where $E_\mathrm{max}$ is defined via Eq.\,(\ref{eq:erg_max}). Figure~\ref{fig:signal} (center) shows that filters with a higher atomic number $Z$ do not necessarily result in a reduced signal because a part of the signal results from the reduced absorption at energies below the respective K-edge. Furthermore, it is shown that a filter material equal to the material of the trace element, here $Z_\mathrm{Au} = 79$, results in the lowest signal. This is because the K-edge of this filter reduces the beam flux mostly at energies slightly above the K-edge where the cross section for the ionization is the largest. The total signal intensity $S$ is plotted as a function of the tube voltage $U_0$ on the right side of the same figure. It is demonstrated that the optimal beam voltage and the filter thickness depend on each other even without considering the background contribution.

\subsection{Background generation}
\label{sec:bkg}
The background spectrum $\mathit{bkg}$ is calculated as the sum of the coherent scattering $\mathit{coh}$ and incoherent scattering $\mathit{inc}$ from the target into the solid angle $\dd \varOmega$ and into the signal interval $E' \pm \dd E'\!/2$
\begin{align}
\label{eq:bkg}
   \mathit{bkg} &= \mathit{coh} + \mathit{inc} \quad \textnormal{in } \left[\frac{\mathrm{ph}}{\kev\textnormal{$\cdot$s$\cdot$g$\cdot$sr}}\right]\\
   \mathit{coh} &= \frac{1}{\rho_{\!A_\mathrm{T}}} \cdot F_\mathrm{F}\cdot\frac{\dd P_\mathrm{coh}}{\dd \varOmega}\\
   \mathit{inc} &= \frac{1}{\rho_{\!A_\mathrm{T}}} \int\limits_{E'}^{E_\mathrm{max}}\!\!\!F_\mathrm{F}\cdot \frac{\dd^2 P_\mathrm{inc}}{\dd \varOmega \, \dd E'}\cdot\mathrm{d}E_0.
\end{align}
Here, $\dd P_\mathrm{coh}/ \dd \varOmega$ is the probability to scatter a photon of energy~$E_0$ coherently into the solid angle $\dd \varOmega$. $\dd^2 P_\mathrm{inc}/ (\dd \varOmega \, \dd E')$ expresses the probability to scatter a photon of energy~$E_0$ incoherently into the signal interval $E' \pm \dd E'\!/2$ and the solid angle $\dd \varOmega$. In the case of coherent scattering there is a one-to-one relation between the beam energy~$E_0$ and the scattered energy $E'$ so that
\begin{equation}
   \frac{\dd P_\mathrm{coh}}{\dd \varOmega} = \frac{\dd(\mu\!/\!\rho)_\mathrm{coh}}{\dd \varOmega} \cdot \rho_{\!A_\mathrm{T}} \quad \textnormal{in } \left[1/\mathrm{sr}\right]\!.
\end{equation}
Using Eq.\,(\ref{eq:coh}), it follows
\begin{equation}
   \label{eq:F_coh}
   \mathit{coh} = F_\mathrm{F} \cdot \frac{\dd \sigma_\mathrm{Th}}{\dd \varOmega} \cdot F^2 \cdot \frac{N_\mathrm{\!A}}{m_\mathrm{a}}
\end{equation}
where Avogadro's constant $N_\mathrm{\!A}$ and the atomic mass of the target element $m_\mathrm{a}$ are used to transform the cross section into a mass attenuation coefficient.
For incoherent scattering Doppler broadening leads to an interval of beam energies that can scatter into the signal interval $E' \pm \dd E'\!/2$, see Fig.\,\ref{fig:doppler}~(right). The double-differential cross section of Eq.\,(\ref{eq:DDCS}) is used to calculate the cross section for each photon with energy $E_0$ to be scattered to the energy \mbox{$E'\pm\dd E'\!/2$}.
With
\begin{equation}
   \frac{\dd^2 P_\mathrm{inc}}{\dd \varOmega \, \dd E'} = \frac{\dd^2(\mu\!/\!\rho)_\mathrm{inc}}{\dd \varOmega \, \dd E'} \cdot \rho_{\!A_\mathrm{T}} \quad \textnormal{in } \left[1/(\kev\textnormal{$\cdot$sr})\right]
\end{equation}
it follows
\begin{equation}
   \label{eq:inc}
   \mathit{inc} = \frac{N_\mathrm{\!A}}{m_\mathrm{a}} \cdot \!\!\int\limits_{E'}^{E_\mathrm{max}}\!\!\!F_\mathrm{F}\,\frac{\dd^2\sigma_\mathrm{inc}}{\dd\varOmega\,\dd E'}\cdot\mathrm{d}E_0.
\end{equation}

\begin{figure}[t]
   \centering
   \includegraphics[width = \linewidth]{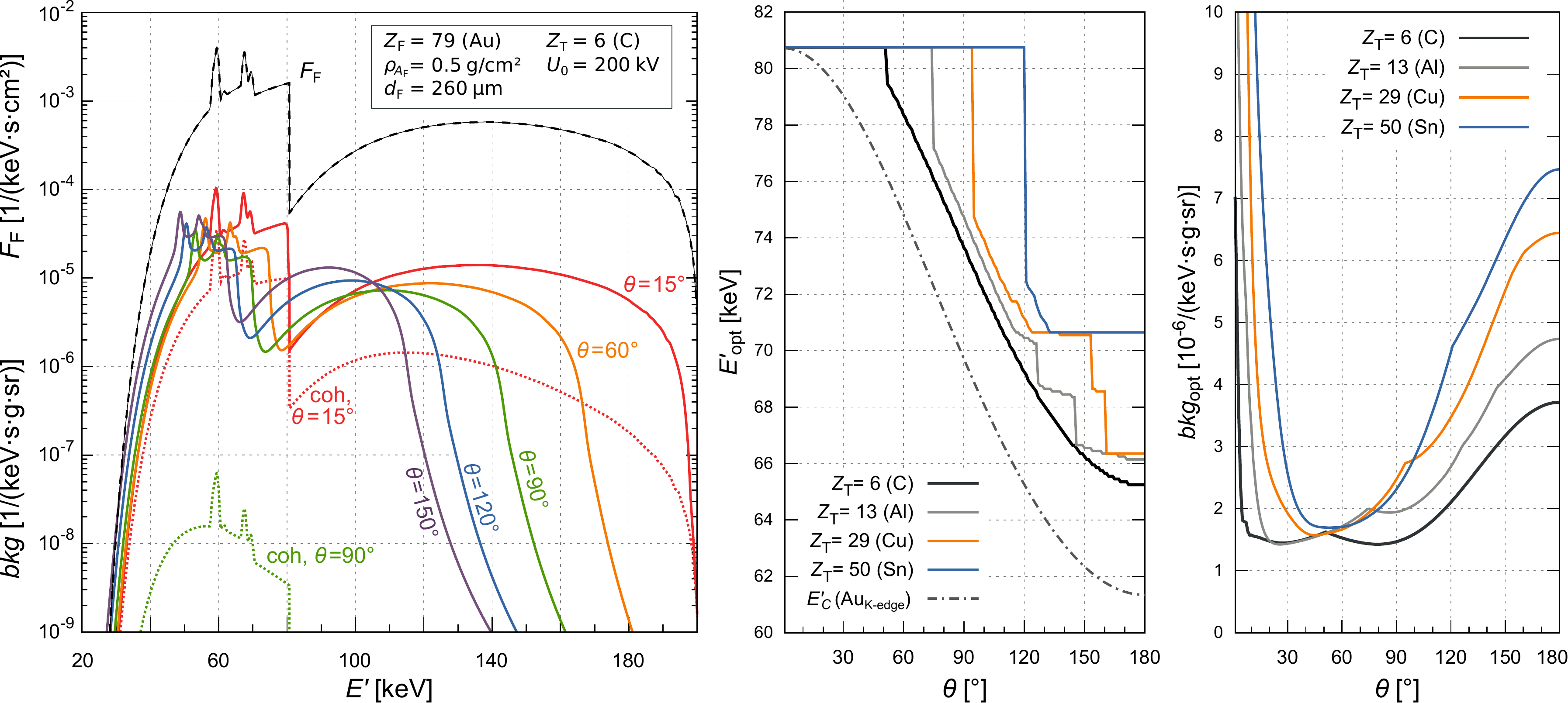}
   \caption{\textbf{Left:} calculation of background spectra for different scattering angles $\theta$. The filtered flux $F_\mathrm{F}$ that serves as input spectrum is shown in dashed lines. The term \textit{bkg} refers to the sum of the coherent and incoherent scattering according to Eq.\,(\ref{eq:bkg}). For demonstration purposes coherent scattering according to Eq.\,(\ref{eq:F_coh}) is shown for \mbox{$\theta=15^\circ$} and \mbox{$\theta=90^\circ$} with dotted lines. \textbf{Center:} energy $E'_\mathrm{opt}$ of the K-edge dip structure after scattering. The Compton energy $E'_\mathrm{c}$ of the K-edge according to Eq.\,(\ref{eq:kk0}) is plotted as dash-dotted line. \textbf{Right:} minimal background at $E'_\mathrm{opt}$.}
   \label{fig:opt_bkg}
\end{figure}

Figure~\ref{fig:opt_bkg} shows calculated background spectra for different scattering angles $\theta$ and an analysis of the minimal background value $\mathit{bkg}_\mathrm{opt}$ and its associated energy $E'_\mathrm{opt}$ for a K-edge dip structure. It is evident that coherent scattering makes a contribution for small scattering angles so that the background spectrum resembles the input spectrum. For increasing scattering angles, coherent scattering becomes negligible and the K-edge dip shifts to lower energies. 
This shift depends on the target material. In addition, the dip position of the K-edge may be modified by additional beam structures like characteristic emission lines that add up via coherent scattering causing a step-like dependency between the energy $E'_\mathrm{opt}$ that minimizes the background and the scattering angle $\theta$, see Fig.\,\ref{fig:opt_bkg} (center). It is important to note that the K-edge dip is less shifted than predicted by the classical Compton shift according to Eq.\,(\ref{eq:kk0}) because of the smearing effect of Doppler broadening. Also note that the energy that minimizes the background for a given scattering angle does not necessarily mean that this scattering angle minimizes the background for this energy. 

The right plot in the same figure shows that the background can be minimized, in this example, for $\theta_\mathrm{opt} \approx 30^\circ$ and $\theta_\mathrm{opt} \approx 80^\circ$ in the case of $Z_\mathrm{T}=6$. Both values are below $\theta=90^\circ$, the angle that minimizes the $\theta$-differential Klein-Nishina cross section given by Eq.\,(\ref{eq:classicCompt}). The presence of two minima can be explained by the sharp K-edge dip structure for $\theta \rightarrow 0^\circ$ in combination with the decreasing cross section for coherent and incoherent scattering for $\theta \rightarrow 90^\circ$.

Even though the presented background spectra show the distribution of the background and its dependency on the scattering angle, they cannot be used to optimize the filter material and thickness because there is no penalty for increasing the filter absorption.

\subsection{Detector response}
Before combining signal and background, the detector response to the signal and the background spectrum must be applied.
Our model for the detector response constitutes of four components: the absorption efficiency of the absorber, the energy resolution of the whole detector system, the Comptonization within the absorber, and the escapes of detector fluorescence lines.
The detailed description of the detector response modeling will be published in an upcoming work, soon. The calculation is summarized in the program \textit{xresp}\,\cite{xresp} and the results are shown in the following sections for the case of a pixelated 625\,$\upmu$m\,x\,625\,$\upmu$m CdTe detector of 1\,mm thickness. Figure~\ref{fig:satbot_blanc_cmp} (left) shows the obtained detector response for $E'=100\,\kev$. \textit{xresp} allows to change the absorber material, its thickness and the pixel pitch and will be publicly available.
In the following, the signal response and the background response of the detector are labeled as $\mathit{sig}^*$ and $\mathit{bkg}^*$, respectively. The energy measured by the detector is labeled as $E_\mathrm{D}$.

\subsubsection*{Comparing model and measurement}
\begin{figure}[ht]
   \centering
   \includegraphics[width = \linewidth]{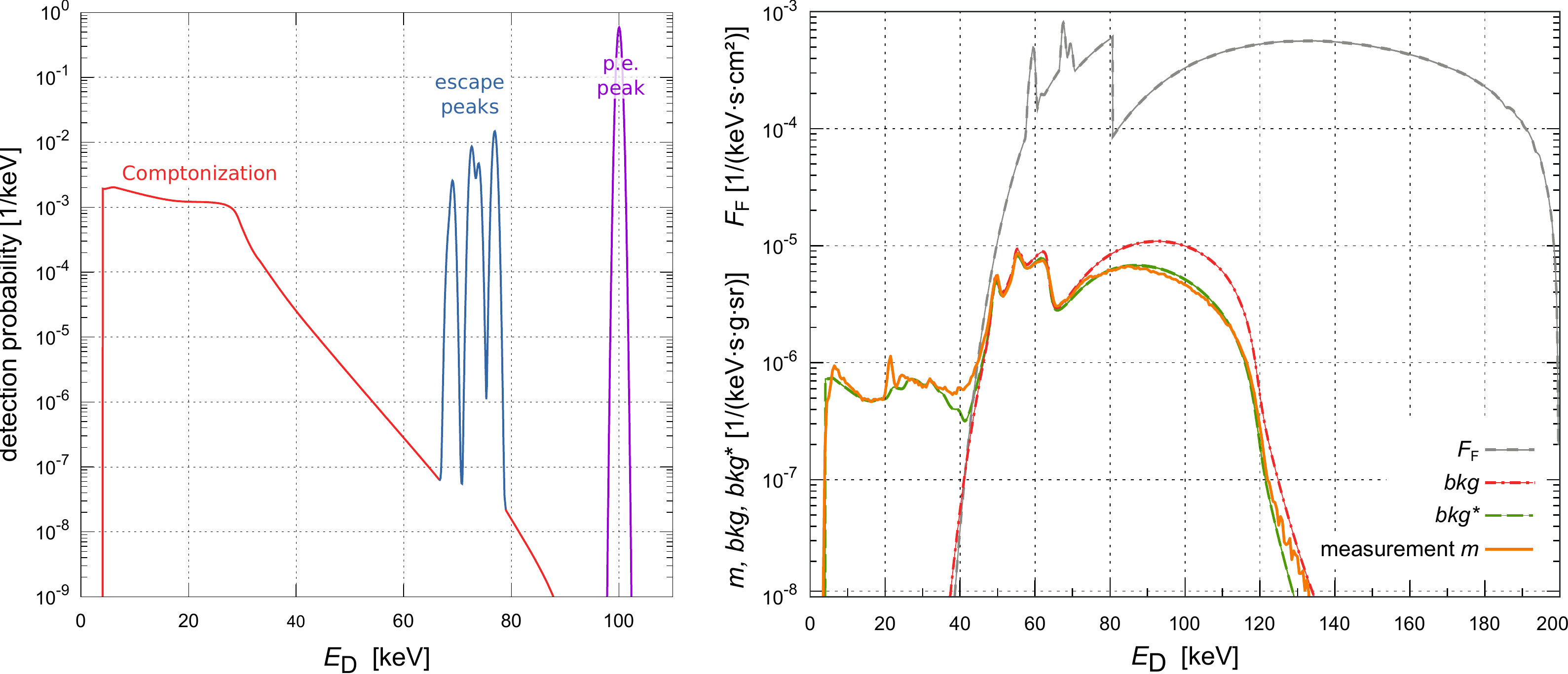}
   \caption{\textbf{Left:} detector response to $E'=100$\,\kev. \textbf{Right:} comparison of the measured background with the background model according to Eq.\,(\ref{eq:bkg}) before ($\mathit{bkg}$) and after ($\mathit{bkg}^*$) applying the detector response. The measurement and the observation used the following parameters: $U_0=200\,$\kv, filter: $150\,\upmu$m Au + 2.0\,mm Cu + 1.5\,mm Al, $\theta = 135^\circ$, \mbox{$Z_\mathrm{T,\,model} = 7$}, while the experiment used as target a water filled plastic tube (\diameter\,11\,mm, outside). Experimental conditions: X-ray tube: Viscom XT9225-DED; collimator: W, 14.5\,mm thick, \diameter\,2.75\,mm, positioned at tube window; distance tube to target: 30\,cm; distance target to detector: 20\,cm. The measured spectrum was recorded with a CdTe based Caliste-HD imaging spectrometer\,\cite{Caliste-HD}.}
   \label{fig:satbot_blanc_cmp}
\end{figure}

Figure~\ref{fig:satbot_blanc_cmp} (right) shows a comparison between a measured background and the computed background model according to Eq.\,(\ref{eq:bkg}) for $U_0=200\,$\kv, \mbox{$\theta = 135^\circ$}, \mbox{$Z_\mathrm{T,\,model} = 7$}, using a $150\,\upmu$m Au + 2.0\,mm Cu + 1.5\,mm Al filter. The excess at $E_\mathrm{D}\approx 21\,$\kev and $E_\mathrm{D}\approx 22$\,\kev is caused by Ag fluorescence within the camera; the small excess at $E_\mathrm{D} \approx 75\,$\kev is caused by Pb fluorescence of the camera shielding; both effects are not included in the model. The deficit at $E_\mathrm{D}\approx40$\,\kev is not fully understood at the moment but including possible charge sharing and charge trapping effects might resolve this difference. The deficit at $E_\mathrm{D} \approx 100$\,\kev and the excess for $E>120$\,\kev might be mainly caused by pile-up effects which are not included in the presented model. Despite these small inaccuracies, the final model matches the observation very well considering the simplicity and the large number of involved models: x-ray tube, filter, scattering in the target, detector response.

\section{Combined signal and background analysis}
Despite its educational purpose, the presented separated analysis of the background and the signal cannot be used to optimize the choice of the filter. A joint analysis for signal and background is presented in this section.
It is shown that optimizations of the signal-to-background ratio ($\mathit{SBR}$) lead to inapplicable results for realistic applications if the signal is allowed to vary. Therefore, another analysis that optimizes the $\mathit{SBR}$ for a given signal strength is presented in Sect.\,\ref{sec:bmin}.

\subsection{\textit{SBR} optimizations for variable signal strengths}

\begin{figure}[ht]
   \centering
   \includegraphics[width = \linewidth]{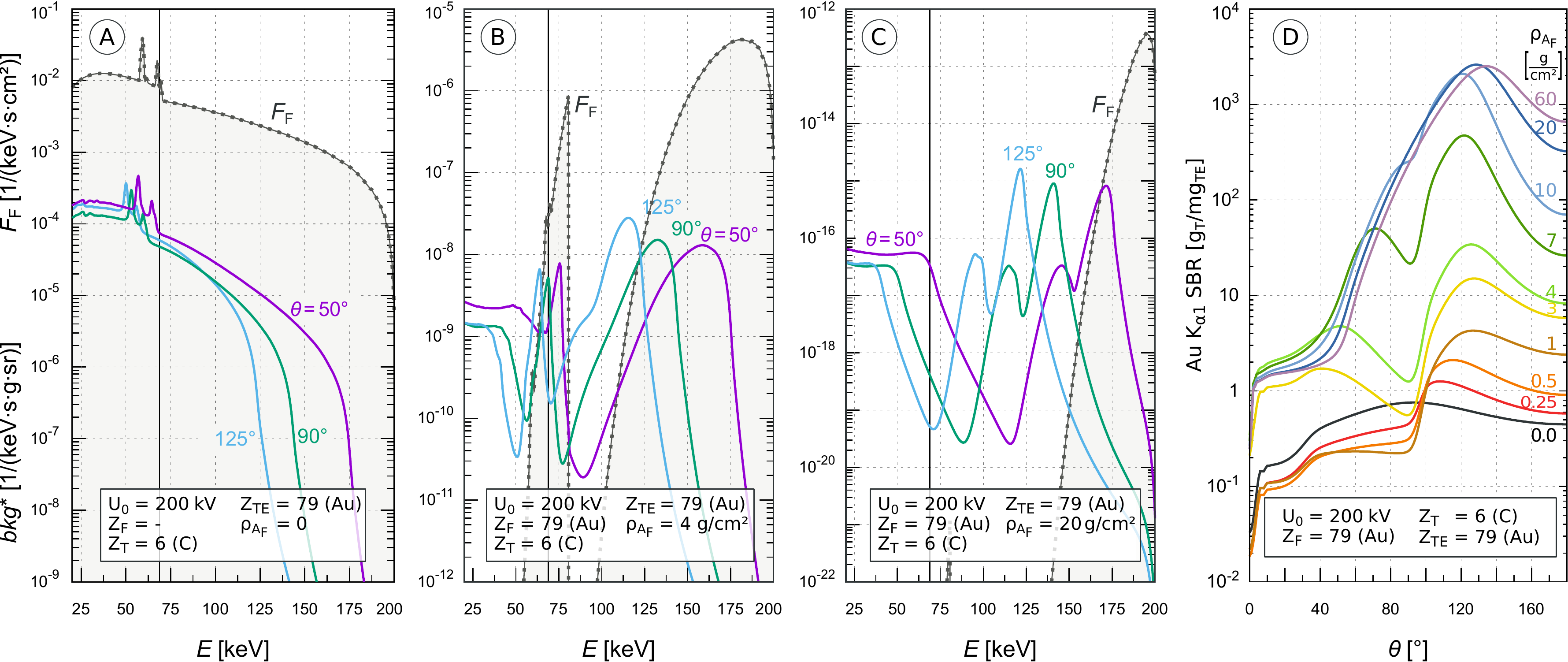}
   \caption{$\mathit{SBR}$ calculations for a Au filter. The three plots on the left (A-C) show the background flux $\mathit{bkg}^*$ for different observation angles $\theta$ and for different filter thicknesses $\rho_\mathrm{\!A}$ in solid lines. For reasons of clarity the signal is not shown  but the position of the Au-K$_{\upalpha 1}$ signal line is indicated with a vertical line. The plot on the right shows the signal-to-background ratio for Au-K$_{\upalpha 1}$ for different filter thicknesses.}
   \label{fig:snr}
\end{figure}

\noindent
Figure~\ref{fig:snr} shows the signal-to-background ratio $\mathit{SBR} = \mathit{sig}^* / \mathit{bkg}^*$ for a Au-K$_{\upalpha 1}$ fluorescence analysis using a Au filter. This example demonstrates three cases:
\begin{itemize}
\setlength\itemsep{0.0em}
\item \textbf{A: unfiltered beam:} the unfiltered beam shows a maximal $\mathit{SBR}$, i.e.~a minimal background, at \mbox{$\theta \approx 90^\circ$} because of a relatively homogeneous input spectrum $F_\mathrm{F}$ and a minimal cross section for incoherent scattering at $\theta \approx 90^\circ$. 
\item \textbf{B: thin filter:} using a gold filter shifts the optimal scattering angle to larger values in order to position the Au-K$_{\upalpha 1}$ line into the reduced background region made by the K-edge of the filter. In addition, a second local minimum of the background is created at lower scattering angles using the left side of the K-edge.
\item \textbf{C: thick filter:} the $\mathit{SBR}$ increases with the filter thickness and reaches a maximal value for \mbox{$\rho_\mathrm{\!A_F}\approx20\,$g/cm$^2$}. The background resembles the detector response for a line energy $E_0 = q_\mathrm{e} \cdot U_0$ that is shifted with the scattering angle according to Eq.\,(\ref{eq:kk0}); see also Fig.\,\ref{fig:satbot_blanc_cmp} (left).
\end{itemize}

Comparing the background of a thin and a thick filter (compare part B and C, see also Fig.\,\ref{fig:opt_bkg}) shows that the K-edge absorption structure is steep for thin filters but becomes a nearly symmetric "V" shape for thicker filters. Therefore, the optimal scattering angle is shifted to larger values for stronger filtering. The optimal area density \mbox{$\rho_\mathrm{\!A_F} \approx 20\,$g/cm$^2$} corresponds to a filter thickness of $d\approx 10\,$mm for a gold filter. This value is so large because the signal results from an integral value of the ionization cross section, see Fig.\,\ref{fig:ion}, while the background results mainly from an integral of the DDCS for incoherent scattering, see Fig.\,\ref{fig:doppler} (right). Comparing the two mentioned figures shows that the DDCS decays much faster for large energies compared to the ionization cross section. As a consequence, strong filtering, which results in a hardening of the X-ray beam, reduces more background than signal.
 
In conclusion, optimizing the $\mathit{SBR}$ by minimizing the dominant background source, i.e.~setting $\theta = 90^\circ$ in order to minimize incoherent scattering, is only valid for an unfiltered and unstructured beam. For thin filters the K-edge structure plays the dominant role if the K-edge is positioned at energies comparable to the fluorescence line energy. For thick filters the K-edge structure is insignificant compared to the nearly mono-energetic beam at the highest possible beam energy, see $F_\mathrm{F}$ in Fig.\,\ref{fig:snr}\,C. In this case the $\mathit{SBR}$ optimization depends mainly on the maximal beam energy, the scattering angle~$\theta$, and the detector response but not on the design of the filter; i.e.~a thick gold filter and a thick tin filter behave similar.

The given analysis does not include considerations about the measurement time which becomes unrealistic long for very thick filters. In addition, the very low spectral flux makes the analysis highly susceptible for background sources that are not included in the model. Therefore, an optimization analysis that focuses on a maximization of the $\mathit{SBR}$  for given signal strengths is presented in the following section.

\subsection{\textit{SBR} optimizations for fixed signal strengths}
\label{sec:bmin}
For practical applications it is useful to define a fixed signal value and optimize the filter material for this signal value. We present an analysis for a reduced signal strength of $f_\mathrm{s}=50$\,\%, 10\,\%, and 1\,\% of the unfiltered signal~$\mathit{sig}_0$. 
The first step is to calculate the area density of the filter that results in the required signal strength $\mathit{sig'}$
\begin{equation}
   \label{eq:signalfix}
   \mathit{sig'}(F_\mathrm{F}) = f_\mathrm{s} \cdot \mathit{sig}_0.
\end{equation}
The signals $\mathit{sig'}$ and $\mathit{sig}_0$ are calculated according to Eq.\,(\ref{eq:sig}) for a given filtered flux $F_\mathrm{F}$ and for an unfiltered flux $F_0$, respectively. Equation~(\ref{eq:signalfix}) is solved numerically for the required area density for each filter material $Z_\mathrm{F}$. In a second step background spectra are calculated for each scattering angle in $1^\circ$ steps. The background value at the bin position of the fluorescence line of interest, here Au-K$_{\upalpha 1}$, is recorded and the minimal background value $\mathit{bkg}_\mathrm{min}$ defines the optimal scattering angle~$\theta_\mathrm{opt}$ and the optimal signal-to-background ratio.

\begin{figure}[ht]
   \centering
   \includegraphics[width = \linewidth]{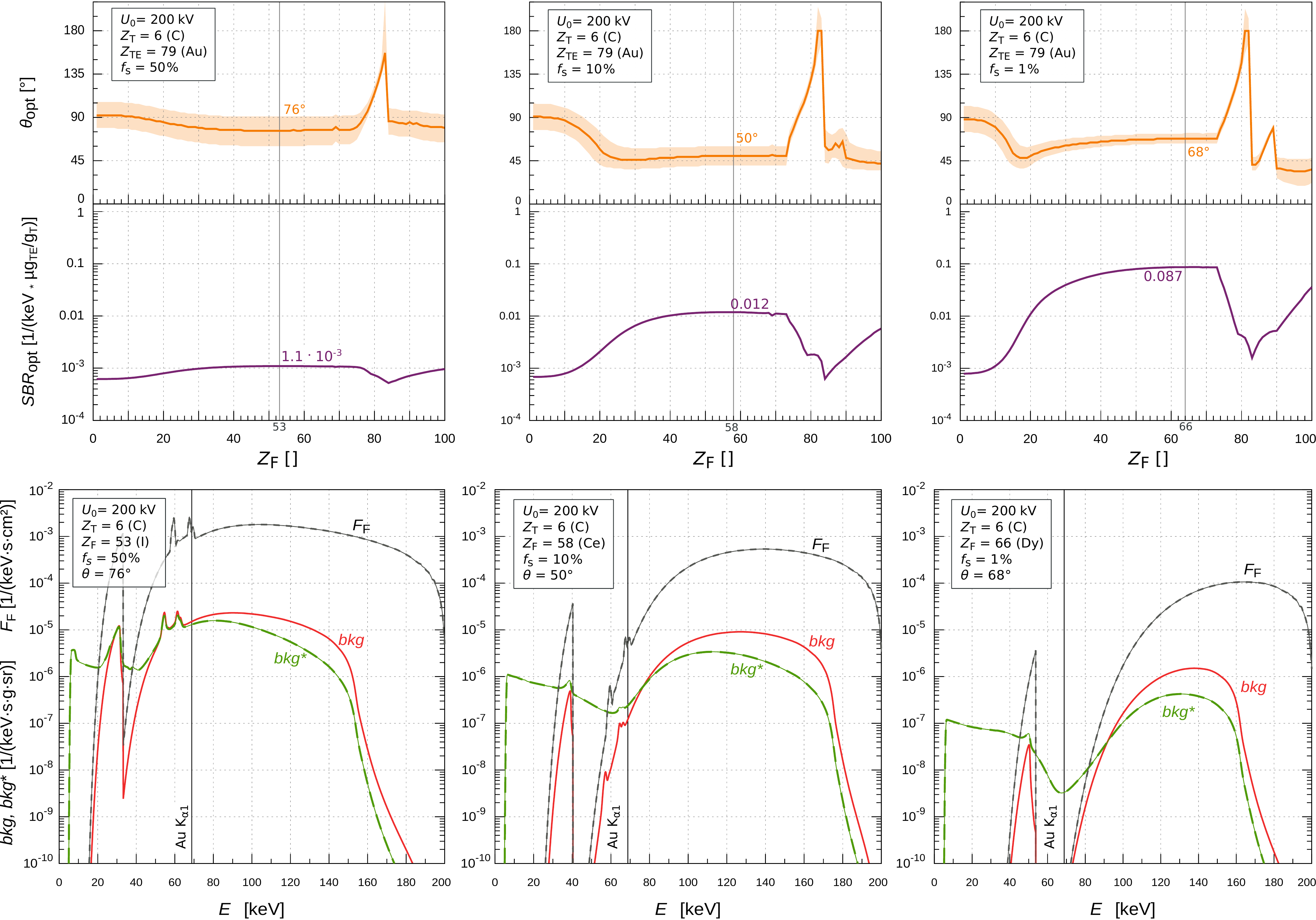}
   \caption{\textbf{Top:} signal-to-background calculations for three fixed signal intensities as function of the filter material $Z_\mathrm{F}$. The filter thickness was chosen for each filter material in a way that the signal has a constant intensity of 50\,\% (left), 10\,\% (center), and 1\,\% (right) of the unfiltered signal. The filter resulting in a maximal $\mathit{SBR}$ is indicated with a vertical line. The values for the $\mathit{SBR}$, the optimal scattering angle, and the optimal filter material $Z_\mathrm{F,opt}$ are also written along this line. The shaded region around $\theta_\mathrm{opt}$ shows an interval of 5\,\% degradation of the $\mathit{SBR}$ with respect to the optimal $\mathit{SBR}$.\newline
   \textbf{Bottom:} filtered flux $F_\mathrm{F}$, the flux that enters the detector $\mathit{bkg}$, and observed background $\mathit{bkg^*}$ for the optimal filter configuration shown above. The fluorescence signals are not shown for reasons of clarity, but the position of the Au-K$_{\upalpha1}$ line is indicated with a vertical line.}
   \label{fig:bmin}
\end{figure}

Figure~\ref{fig:bmin} shows the  $\mathit{SBR}$ and the optimal scattering angle $\theta_\mathrm{opt}$ for all filter elements $Z_\mathrm{F}$ for the case $U_0=200\,$\kv, $Z_\mathrm{T}=6$, and $Z_\mathrm{TE}=79$. An interval of 5\,\% degradation of the optimal $\mathit{SBR}$ value around $\theta_\mathrm{opt}$ indicates the influence of the scattering angle to the $\mathit{SBR}$. It can be seen:
\begin{itemize}
\setlength\itemsep{-0.3em}
\item The $\mathit{SBR}$ is increased by a factor of $\sim$80 going from $f_\mathrm{s} = 50\,\%$ to  $f_\mathrm{s} = 1\,\%$.
\item The maximum of the $\mathit{SBR}$ is relatively wide. The maximal $\mathit{SBR}$ defines $Z_\mathrm{F,opt}$ but the filter material can be chosen out of a larger range of available filters without a big increase of the background. From here on, we call all filters with a $\mathit{SBR}$ close to the optimal filter a \textit{matched filter}. In the examples shown in Fig.\,\ref{fig:bmin} the interval size for matched filters is $\Delta Z_\mathrm{F} \ge 20$ accepting a maximal background increase of 4\,\%.
\item The differences between a matched filter and an unmatched filter scales with the filter thickness. For a 1\,\% signal intensity the $\mathit{SBR}$ of a matched filter is 100 times higher compared to the worst filter selection.
\item The interval size of matched filters decreases for stronger filtering.
\item The optimal filter material is changing with the relative signal strength: smaller values of $f_\mathrm{s}$ result in filter with larger atomic numbers $Z_\mathrm{F}$.
\item $\theta_\mathrm{opt}$ is changing with $f_\mathrm{s}$ without a recognizable rule.
\item The 5\,\% degradation interval is decreasing, i.e.~the sensitivity of $\theta_\mathrm{opt}$ is increasing for thicker filters.
\item Only strong filtering with $f_\mathrm{s}=1\,\%$ forms a clear K-edge dip at the position of the fluorescence line.
\item Using a strong filter requires an accurate detector response model as the background at Au-K$_{\upalpha 1}$ is mainly generated by the detector response.
\end{itemize}

\begin{figure}[hbt]
   \centering
   \includegraphics[width = \linewidth]{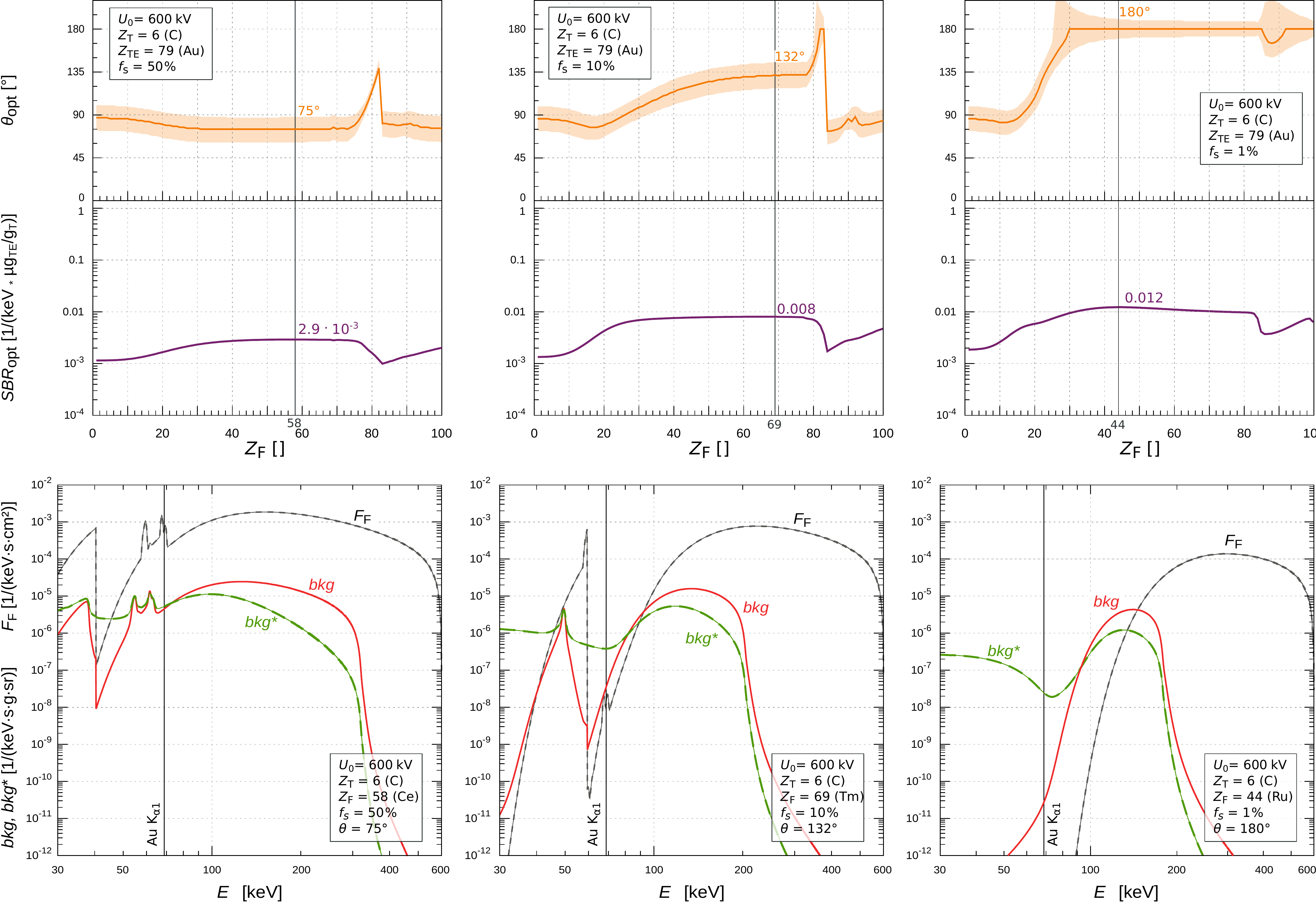}
   \caption{same as Fig.\,\ref{fig:bmin} but with $U_0=600\,$\kv.}
   \label{fig:bmin2}
\end{figure}

Figure~\ref{fig:bmin2} shows the same analysis for $U_0 = 600\,$\kv. Here, the optimal scattering angle is considerably larger, especially for strong filtering.
Note also that the filter for $f_\mathrm{s}= 1\,\%$ clearly represents the case for a thick filter.
Another subtlety here is that the fluorescence line cannot be positioned in the dip of the background even for $\theta_\mathrm{opt} = 180^\circ$.
In this case the background is mainly defined by the detector response and not by the filter characteristics which causes an $\mathit{SBR}$ increase by a factor of $\sim$4 going from $f_\mathrm{s} = 50\,\%$ to  $f_\mathrm{s} = 1\,\%$ which is 20 times less compared to the case $U_0 = 200\,$kV.

\subsection{Choice of the tube voltage}
\begin{figure}[tbh]
   \centering
   \includegraphics[width = 0.90\linewidth]{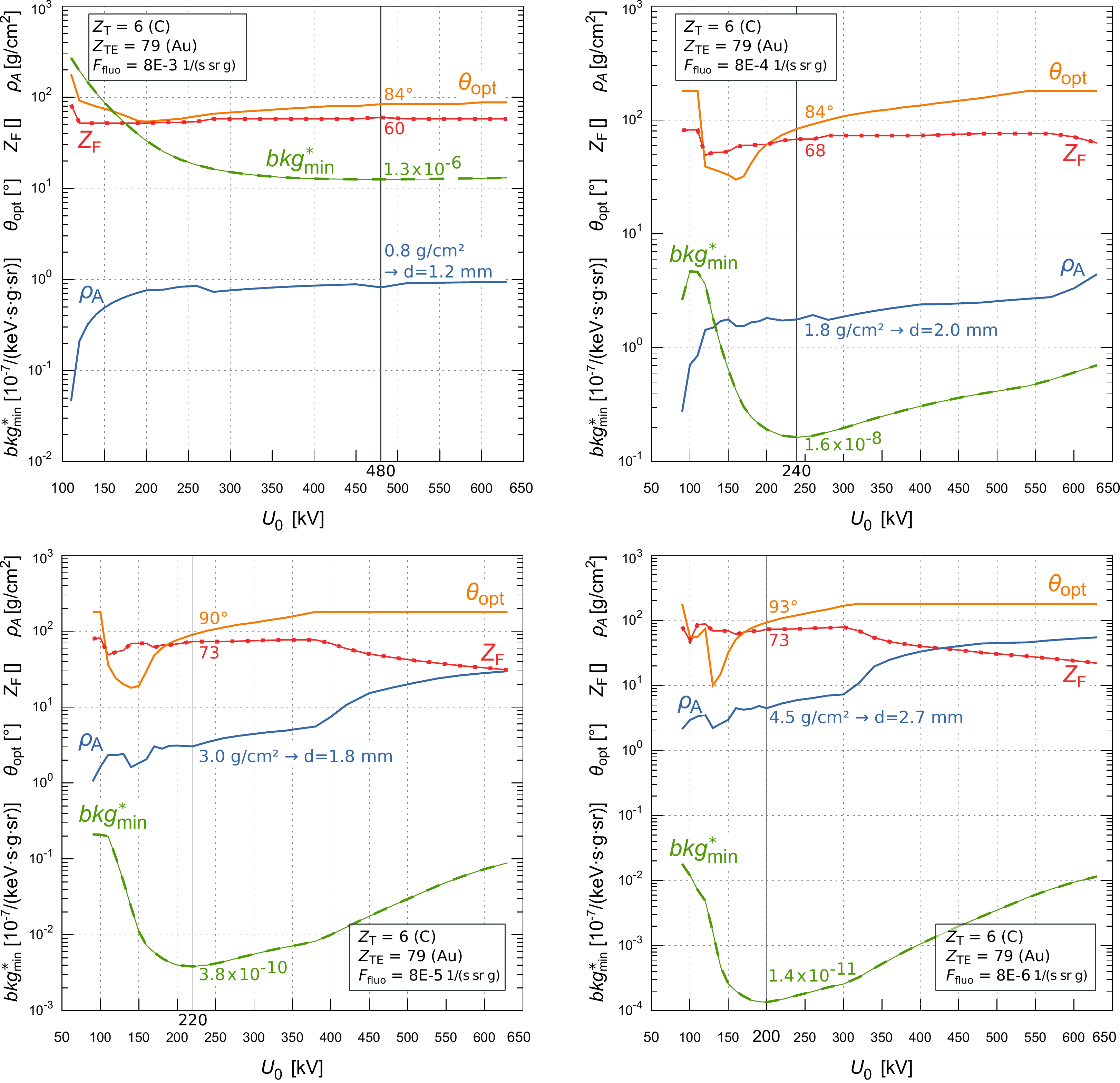}
   \caption{Optimal tube voltage $U_0$ for Au-fluorescence in carbon for different fluorescence intensities $F_\mathrm{fluo}$. The vertical line is positioned at the minimum of $\mathit{bkg}^*_\mathrm{min}$.}
   \label{fig:Uopt}
\end{figure}

The analysis in the last section already demonstrated a dependency between the tube voltage $U_0$ and the filter optimization. To quantify this dependency the analysis for $\mathit{bkg}_\mathrm{min}$, $\theta_\mathrm{opt}$, and $Z_\mathrm{F,opt}$ were redone for different tube voltages in the range $90\,\mathrm{\kv} \le U_0 \le 630$\,\kv. As the signal $\mathit{sig}_0$ depends on $U_0$ a background analysis for a relative signal strength, as done in the previous section, is not reasonable for different $U_0$. Therefore, the analysis uses fixed fluorescence intensities $F_\mathrm{fluo}\,[\textnormal{1/(s$\cdot$g$\cdot$sr)}] \in \{8\cdot\!10^{-3},8\cdot\!10^{-4},8\cdot\!10^{-5},8\cdot\!10^{-6}\}$. As a consequence of the fixed signal the optimization of the $\mathit{SBR}$ becomes a minimization of the background.
Figure~\ref{fig:Uopt} shows the results for Au-fluorescence in carbon:
\begin{itemize}
   \item The optimal tube voltage $U_\mathrm{0,opt}$ that results in a minimal background is decreasing from 480\,\kv to 240\,\kv, 220\,\kv, and 200\,\kv for decreasing fluorescence intensities \mbox{$F_\mathrm{fluo}$}.
   \item The interval of tube voltages that results in a low background is becoming smaller for stronger filtering.
   \item The optimal scattering angles for $U_\mathrm{0,opt}$ are close to $\theta = 90^\circ$ and are slightly increasing for stronger filtering.
   \item The optimal filter material $Z_\mathrm{F,opt}$ is slightly increasing for stronger filtering.
\end{itemize}

\section{Summarizing discussion}
In addition to the results and their discussions in the previous sections we present an extended discussion of the main results in this section.
Even though some rules for the optimal tube voltage, the optimal filter material and the optimal observation angle are stated for the given examples, a repetition of this analysis should be done for each experiment different to the presented examples.
A typical workflow for an XRF filter optimization with the corresponding \textit{xfilter} commands and a discussion of their results is shortly presented in the annex (Sect.\,\ref{sec:annex}) for the example of gold fluorescence in human tissue.

If the tube voltage can be chosen an optimization of $U_0$ is equally important as an optimization of the filter material; a poorly chosen tube voltage can degrade the $\mathit{SBR}$ by an order of magnitude or more.
Compared to $U_0$ and $Z_\mathrm{F}$ the observation angle seems to be less sensitive to the signal-to-background ratio.
The widely used rule of $\theta_\mathrm{opt} = 90^\circ$ is not always true but a good choice without any prior knowledge. In the demonstrated examples the rule is valid if the tube voltage is close to the optimal voltage: $U_0 \approx U_\mathrm{0,opt}$. For smaller tube voltages the optimal scattering angle is decreasing and for larger tube voltages the optimal scattering angle is increasing.
$\theta = 90^\circ$ is also a good choice for an unfiltered beam if the spectral region of interest is relatively unstructured. Thick filters result in a quasi monoenergetic beam and the optimal scattering angle can be calculated, knowing the detector response. 

In all other cases, i.e.~for thin and medium thick filters, a specific filter analysis is required. The thicker the filter, the more pronounced is a v-shaped K-edge absorption structure. The strong dip of this structure results in a more sensitive relation between the optimal scatting angle and the maximal $\mathit{SBR}$. Thinner filters are not dominated by the K-edge absorption, but by the initial shape of the X-ray tube which is, except for the characteristic radiation, relatively smooth. To keep the scattering angle close to the optimal value becomes therefore, in general, less important.

\subsection*{Filter thickness}
The program collection \textit{xfilter} can be used to obtain the optimal filter material and the optimal observation angle. The tube voltage can be also optimized or set fixed. The filter thickness depends on other constraints, like the measurement time, dose deposition, beam intensity, or noise considerations and must be chosen for each setup individually. In general the filter thickness should be chosen as thick as possible and as thin as necessary. Even though the optimization of the filter thickness is not directly addressed by \textit{xfilter}, it can be used to calculate the necessary thickness for a given signal strength.

\subsection*{Filter fluorescence radiation}
Fluorescence radiation emitted by the filter is not considered within the presented calculations and can deteriorate the performance of the setup. The annex shows an example how a filter can be chosen different to the optimal filter material with only a small increase in background. In addition, a proper shielded and collimated filter, i.e.~a strong collimator behind the filter helps to minimize the contribution of the filter fluorescence. If this is not sufficient, an additional filter can be used to filter the fluorescence lines of the primary filter. The additional filter should be kept thin, as scattering effects blur any spectroscopic structure produced by the primary filter.

\subsection*{Poisson noise}
This work presents a filter optimization assuming a counting statistics that is high enough to ignore Poisson noise. Dependent on the measurement time, beam intensity, filter thickness, detector size, distance between tube and target and distance between target and detector, the counting statistics can become small enough that Poisson noise needs to be considered. These considerations will mainly affect the filter thickness but not the filter material, or the observation angle, or the tube voltage. Therefore, noise considerations can be made independently from the presented optimizations in a subsequent step.

\subsection*{Trouble shooting}
If the measured background dissents from the calculated value a careful background analysis have to be conducted. In particular, the camera shielding, the detector response model, and the proper installation of beam and camera collimators should be checked.

\subsection*{Summary}
The presented program collection \textit{xfilter} allows to optimize EDXRF measurements that are using X-ray tubes with $20\,\mathrm{\kv} \le U_0 \le 640\,\mathrm{\,\kv}$. The optimization derives from a filter based background minimization that can be applied for any combination of target materials and trace elements resulting in optimal values for the tube voltage, the filter material, and the observation angle. The background calculation for a water filled plastic tube showed excellent agreement with the observation, demonstrating a precise modeling of the X-ray tube, the filter, the background diffusion in the target, and the detector response. The modeling of the detector response is out of the scope of this work and must be given as external input to \textit{xfilter}. A generalized detector response generator\,\cite{xresp} which was already used for this work will be presented in an upcoming publication.

\subsection*{Acknowledgement}
We acknowledge the financial support of DRF-impulsion interdisciplinary program of CEA, the French Alternative Energies and Atomic Energy Commission.
The authors wish to thank Sylvie Chevillard and Romain Grall of CEA/DRF/IRCM, Adrien Stolidi, Caroline Vienne, and Hermine Lemaire of CEA/DRT/LIST/DISC and C\'ecile Sicard and Emilie Brun of Universit\'e Paris Sud for fruitful discussions in the frame of the DRF-impulsion SATBOT project, motivating the present research on X-ray filter optimization for EDXRF analysis.

\section{Annex}
\label{sec:annex}
\begin{figure}[phtb]
   \centering
   \includegraphics[width = 0.86\linewidth]{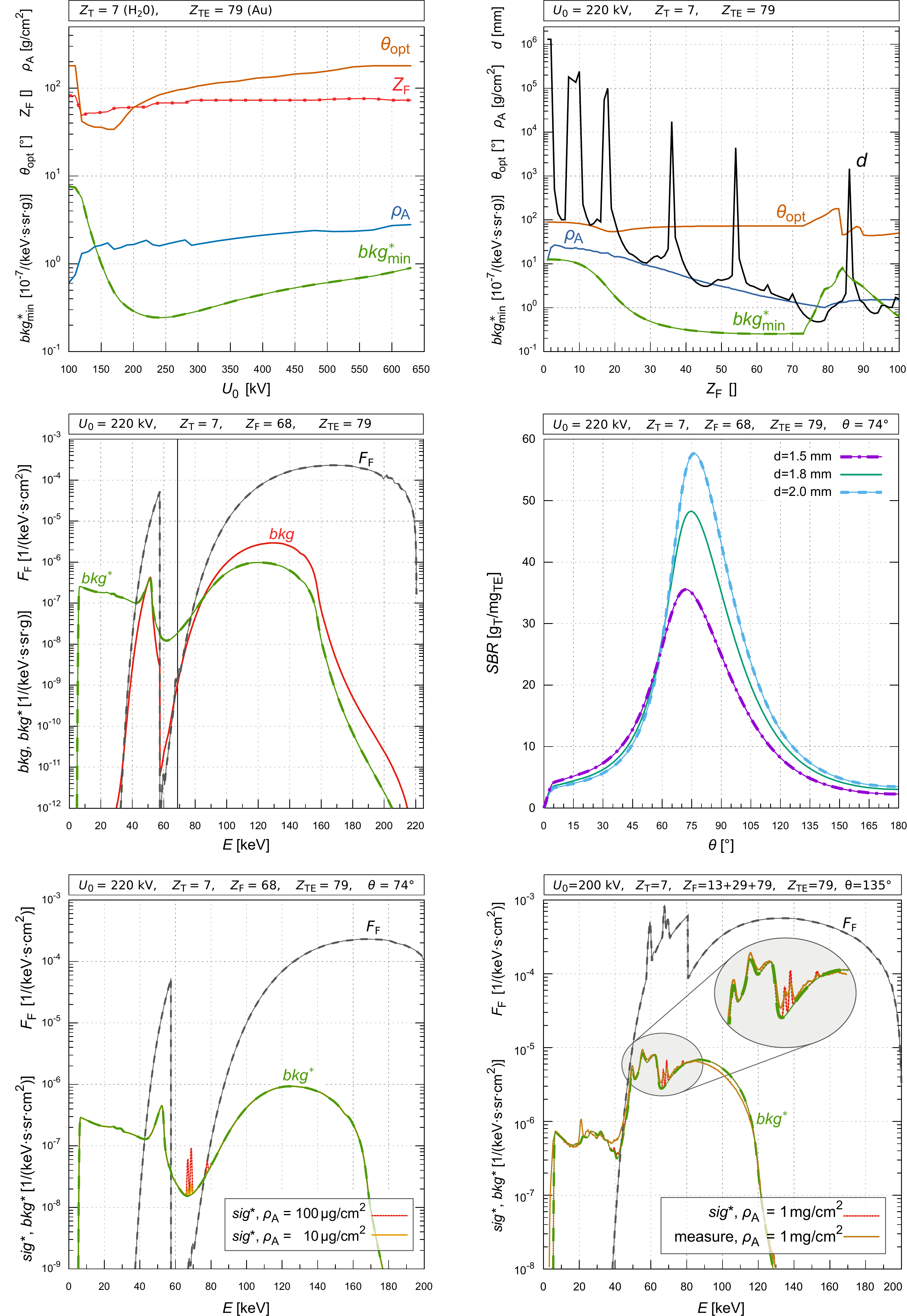}
   \caption{Calculation results for the minimal background (top, left), optimal filter material (top, right), background spectrum (center, left), signal-to-background ratio (center, right), and final spectrum (bottom, left) for the example of Au-K$_{\upalpha 1}$ XRF in human tissue  ($Z_\mathrm{T}=7$). A measurement with different setup parameters is shown bottom, right.}
   \label{fig:end}
\end{figure}

The following workflow shows all steps of an XRF filter optimization using \textit{xfilter}:
\begin{enumerate}
   \item Use \textit{xfilterVolt} for a tube voltage analysis like the one shown in Fig.\,\ref{fig:Uopt}. Find $U_\mathrm{0,opt}$ and the corresponding values for $Z_\mathrm{F,opt}$, $\theta_\mathrm{opt}$, and $\rho_{\!A}$. If $U_0$ is given by other constraints like the maximal accessible tube voltage use this value and find the corresponding values for $Z_\mathrm{F,opt}$, $\theta_\mathrm{opt}$, and $\rho_{\!A}$.
   \item Use \textit{xfilterFilter} for a filter analysis like the one shown in Fig.\,\ref{fig:bmin} (top). This allows to see how much the background changes for filter materials different to $Z_\mathrm{F,opt}$. 
   \item Use \textit{xfilterSpecB} for a spectral analysis like the one shown in Fig.\,\ref{fig:bmin} (bottom). This allows to get the background spectrum and to identify possible spectral features that make this choice less favorable. \mbox{This step} might be repeated for other filter materials until a good filter is found.
   \item Use \textit{xfilterSBR} for a signal-to-background analysis like the one shown in Fig.\,\ref{fig:snr} (right). This step allows to see the consequences for different filter thicknesses. Choose final filter thickness and scattering angle. 
   \item  Use \textit{xfilterSpecS} to get the final expected spectrum, see Fig.\ref{fig:end} (bottom, center and right).
\end{enumerate}

Figure~\ref{fig:end} shows the results for a sequence of calculations for the case of gold fluorescence in human tissue. The atomic number of the target is approximated with $Z_\mathrm{T} \!=\! 7$. The fluorescence intensity is set to \mbox{$F_\mathrm{fluo} = 0.001$\,1/(s$\cdot$g$\cdot$sr)}. The area density for the trace elements is set to $\rho_\mathrm{\!A,TE} = 10\,\upmu$g/cm$^2$ and $\rho_\mathrm{\!A,TE} = 100\,\upmu$g/cm$^2$ and for the target to $\rho_\mathrm{\!A,T} = 1\,$g/cm$^2$. In the following, first the used command for the calculation (with computation time and memory usage) is shown followed by a short discussion. A description of the program parameters can be found in reference\,\cite{xfilter}.

\vspace{2mm}
\noindent
\textbf{Command} (70\,min 09\,s; 4.1\,GB and 126\,min 48\,s; 6.5\,GB):
\vspace{-2mm}
\begin{verbatim}
./xfilterVolt 100 300 10 7 2 79 1 0.001 CdTeResp.dat 
./xfilterVolt 330 630 30 7 5 79 1 0.001 CdTeResp.dat 
\end{verbatim}
\vspace{-2mm}
Here, the calculation is split in two calculations with a higher resolution of $\Delta U_0=10\,\kv$ and $\Delta \theta=2^\circ$ for $100 \le U_0\,[\mathrm{\kv}] \le 300$.
Figure~\ref{fig:end} (top, left) shows that the optimal tube voltage is at $U_0 = 240\,$\kv. But it also shows that the background is only minimally increasing for slightly different tube voltages. Because of the maximal voltage of the used X-ray tube we choose $U_0=220\,$\kv accepting a 3.3\,\% increase of background.

\vspace{2mm}
\noindent 
\textbf{Command} (11\,min 30\,s; 4.4\,GB):
\vspace{-2mm}
\begin{verbatim}
./xfilterFilter 220 7 1 79 1 0.001 CdTeResp.dat
\end{verbatim}
\vspace{-2mm}
Figure~\ref{fig:end} (top, right) shows that the optimal filter material for $U_0 = 220\,$\kv is $Z_\mathrm{F,opt} = 66$. The figure also shows that all filter materials with $50 \le Z_\mathrm{F} \le 72$ work well for us. Arguments of availability, cost, chemical stability, toxicity or filter fluorescence might lead to choose the filter material different to $Z_\mathrm{F,opt}$. Choosing $Z_\mathrm{F}=68$ (0.06\,\% increase in background) leads to $\theta_\mathrm{opt}=74^\circ$ and $d \approx 1.8\,$mm.

\vspace{2mm}
\noindent
\textbf{Command} (0\,min 36\,s; 0.2\,GB):
\vspace{-2mm}
\begin{verbatim}
./xfilterSpecB 220 68 -1.8 7 85 CdTeResp.dat 
\end{verbatim}
\vspace{-2mm}
The calculated background spectrum is shown in Fig.\,\ref{fig:end} (center, left).
The obtained spectrum can be used to check other constraints like nearby peaks, the background at other energies, or the total dose.

\vspace{2mm}
\noindent
\textbf{Command} ($3 \times 2$\,min 51\,s; 0.2\,GB):
\vspace{-2mm}
\begin{verbatim}
./xfilterSBR 220 68 -1.5 7 1 79 1 CdTeResp.dat
./xfilterSBR 220 68 -1.8 7 1 79 1 CdTeResp.dat
./xfilterSBR 220 68 -2.0 7 1 79 1 CdTeResp.dat
\end{verbatim}
\vspace{-2mm}
The signal-to-background calculation is done for similar filter thicknesses to the chosen one in order to see how the optimal scattering angle varies if the filter thickness is changed, see Fig.\,\ref{fig:end} (center, right). Be aware that the values of the calculation have units of 1 but the plot uses \textit{gram of target element per milligram of trace element}.

\vspace{2mm}
\noindent
\textbf{Command} ($2 \times 0$\,min 36\,s; 0.2\,GB):
\vspace{-2mm}
\begin{verbatim}
./xfilterSpecS 220 68 -1.8 7 1 74 79 1e-4 CdTeResp.dat 
./xfilterSpecS 220 68 -1.8 7 1 74 79 1e-5 CdTeResp.dat 
\end{verbatim}
\vspace{-2mm}
Finally, the expected XRF spectrum can be calculated. The signal calculation is done for 100\,$\upmu$g/cm$^2$ and for 10\,$\upmu\mathrm{g/cm}^2$, see Fig.\,\ref{fig:end} (bottom, left).

\subsection{Comparing to measurement}
A measurement that did not use the presented framework to optimize the EDXRF setup, but that was optimized by an experimental approach is shown in Fig.\,\ref{fig:end} (bottom, right). A different tube voltage, filter material, and observation angle leads to a background value that is two orders of magnitude larger compared to the optimal filter solution shown in the same figure on the bottom, left.

\bibliographystyle{elsarticle-num} 
\bibliography{lit}

\end{document}